\documentclass[twocolumn,aps,floatfix,superscriptaddress,longbibliography]{revtex4-2}
\usepackage{amsmath,amssymb,eucal,graphicx,float,epstopdf,xparse,verbatim}
\usepackage{epsfig,subfigure}
\usepackage{booktabs,makecell,diagbox}
\usepackage[utf8]{inputenc}

\usepackage[colorlinks=true, urlcolor=blue, anchorcolor=blue, citecolor=blue,filecolor=blue,linkcolor=blue,menucolor=blue]{hyperref}

\begin{document}
\title{Leaves of preferential attachment trees}

\author{Harrison Hartle}
\email{hhartle@santafe.edu}
\affiliation{Santa Fe Institute, Santa Fe, New Mexico 87501, USA}

\author{P. L. Krapivsky}
\email{pkrapivsky@gmail.com}
\affiliation{Santa Fe Institute, Santa Fe, New Mexico 87501, USA}
\affiliation{Department of Physics, Boston University, Boston, Massachusetts 02215, USA}

\begin{abstract}
We provide a local probabilistic description of the limiting statistics of large preferential attachment trees in terms of the ordinary degree (number of neighbors) but augmented with information on leafdegree (number of neighbors that are leaves). The full description is the joint degree-leafdegree distribution $n_{k,\ell}$, which we derive from its associated multivariate generating function. From $n_{k,\ell}$ we obtain the leafdegree distribution, $m_{\ell}$, as well as the fraction of vertices that are protected (nonleaves with leafdegree zero) as a function of degree, $n_{k,0}$, among numerous other results. We also examine fluctuations and concentration of joint degree-leafdegree empirical counts $N_{k,\ell}$. Although our main findings pertain to the preferential attachment tree, the approach we present is highly generalizable and can characterize numerous existing models, in addition to facilitating the development of tractable new models. We further demonstrate the approach by analyzing $n_{k,\ell}$ in two other models: the random recursive tree, and a redirection-based model.
\end{abstract}

\maketitle

\section{Introduction}

Studies of real and random graphs often use degree as a key quantifier of vertex centrality \cite{barabasi2013network}. The degree distribution $\{n_k\}_{k\ge 1}$ determines numerous properties of graphs \cite{molloyreed} and is widely examined in real networks \cite{Newman}, which often exhibit highly right-skewed degree statistics \cite{voitalov2019scale}, including cases in which a powerlaw provides a useful description \cite{broido2019scale}. The presence of powerlaws has been estimated in many scenarios and explained by many idealized mechanisms \cite{newman2005power,clauset2009power}. The paradigmatic mechanistic models of powerlaw-tailed networks are growth models such as preferential attachment (PA). The simplest form of PA consists of growing random tree with degree-proportionate attachment probability \cite{Simon55,barabasi99}. PA models constitute a widely investigated subject both in random graph theory and network science \cite{DM03,Hofstad,Frieze}. Early findings include the exact degree distribution \cite{KR00,Sergey00} and associated self-averaging \cite{bollobas2003mathematical}, the vanishing of the epidemic threshold \cite{PhysRevE.63.066117}, divergence of Ising critical temperature \cite{BIANCONI2002166}, and characterization of the diameter \cite{bollobas2004diameter}, degree correlations \cite{KR01,nikoloski2005degree}, and depth-stratified degree distribution \cite{ben2009stratification}.

Despite the detailed characterization of degree statistics in the PA tree, a closely related aspect has gone uncharacterized: the statistics of leaves (degree one vertices) \cite{Diestel}, and in particular, the leafdegree distribution, with the leafdegree of a vertex being the number of leaves that it is attached to \cite{LIN202597}. The leafdegree distribution has not been examined in almost any models nor real graph data, in contrast with the degree distribution, which is ubiquitously examined \cite{DM22}. The importance of leafdegree distribution is motivated by the abundance and distinguished role of leaves in trees \cite{Drmota} and other sparse graphs \cite{https://doi.org/10.1002/rsa.20334}. Leaves are the outermost boundary points of graphs and have natural interpretations in real systems. For instance, leaves describe terminal units in transport networks \cite{west}, observed or extinct species in phylogenetic trees \cite{Bachmaier2005}, and non-transmitting infected individuals in contagion trees \cite{infection2014}. Leaves also lead to the definition of rank (distance to nearest leaf) \cite{Pittel17}, equaling $0$ for leaves, $1$ for neighbors of leaves, and $\ge 2$ for the remaining fraction $p$ consisting of protected vertices; see \cite{10.1214/ECP.v19-3048,mahmoud2015asymptotic,Bona14,Prodinger12,Devroye14}.

Recent work has shown that some models admit closed recurrences for the leafdegree distribution $\{m_\ell\}_{\ell\ge 0}$ \cite{hartle2025statistics}, with $m_\ell$ denoting the limiting fraction of vertices with leafdegree $\ell$. Namely, for the random recursive tree (RRT), for a leafdegree-based preferential attachment model, and for unlabeled statistical variants of both \cite{hartle2025growing}. However, a closed recurrence for $\{m_\ell\}_{\ell\ge 0}$ does not exist for the original PA tree, since the degree, and not just the leafdegree, impacts how leaf statistics evolve during growth. This degree dependence enabled the closed recurrence for the powerlaw-tailed degree distribution in PA \cite{Simon55,barabasi99}, which subsequently contributed to numerous developments in the foundations of network science \cite{DM22,barabasi2013network}. A natural question is whether the leafdegree statistics are also heavy-tailed. 

The absence of a closed recurrence for $\{m_{\ell}\}_{\ell\ge 0}$ in PA calls for a wider and more comprehensive description. To recover and extend both the degree distribution and leafdegree distribution, we consider the joint distribution $n_{k,\ell}$ of degree and leafdegree. We find a closed and tractable two-index recursion for PA that encodes all values of $n_{k,\ell}$. Obtaining $n_{k,\ell}$ then yields not only $m_{\ell}=\sum_{k=\ell+1}^{\infty}n_{k,\ell}$, which we find indeed is heavy-tailed ($m_\ell\sim \ell^{-3}$), but also allows calculation of other quantities of interest. For instance, we obtain the protected fraction $p=0.039\, 447\, \ldots$, and the degree distribution among protected vertices; $n_{k,0}\simeq k^{-3}2^{-k}$. Analogous calculations may be performed for related tree growth models; see Appendices~\ref{rrt},~\ref{app:crmodel} and Sec.~\ref{discussion}. We focus on PA trees, but our methods can be extended to more general sparse random graphs; this work brings into focus a family of outstanding theoretical questions and domains for further investigation.

This article is organized as follows. In Sec.~\ref{results}, we describe our approach and present our main findings. We focus entirely on PA in the main text, but provide a detailed analysis of the RRT in Appendix~\ref{rrt} and a redirection-based model in Appendix~\ref{app:crmodel}. In Secs.~\ref{pa}--\ref{fluctuations}, we derive our main results for PA trees. Specifically, in Sec.~\ref{fluctuations}, we characterize random fluctuations, demonstrating extensivity and asymptotic self-averaging. This ensures that nonrandom limiting fractions $n_{k,\ell}$ exist for the PA tree, which we obtain in Sec.~\ref{pa} via generating function. Some technical derivations of the results from Sec.~\ref{fluctuations} are relegated to Appendices \ref{ap:Gauss}--\ref{ap:cum}. In Sec.~\ref{discussion}, we briefly conclude.

\section{Approach and main results}
\label{results}

We consider an $N$-vertex growing random tree $G_N$. Its vertices $j=1,\ldots,N$ are labeled sequentially by arrival order. The number of neighbors or degree of vertex $j$ is denoted $k_j$. Leaves are vertices for which $k_j=1$. The number of neighbors of $j$ that are leaves is denoted $\ell_j$ and called the leafdegree of $j$. The inequality $\ell_j<k_j$ holds for all vertices $j$ in all trees $G_N$ unless $j$ is the central vertex in a star graph (in which case $k_j=\ell_j=N-1$). The number $N_{k,\ell}$ of vertices with $k_j=k$ and $\ell_j=\ell$ is a random quantity. Extensivity and asymptotic self-averaging imply convergence in probability to nonrandom limiting fractions $\frac{1}{N}N_{k,\ell}\rightarrow n_{k,\ell}$ as $N\rightarrow\infty$, justifying our analysis. The ordinary degree and leafdegree distributions are determined by the joint distribution:
\begin{equation}\begin{aligned}
n_k=\sum_{\ell=0}^{k-1}n_{k,\ell}, \qquad m_\ell=\sum_{k=\ell+1}^{\infty}n_{k,\ell}. 
\end{aligned}\end{equation}
In addition to the normalization condition, $\sum_{0\le \ell< k<\infty}n_{k,\ell}=1$, the fractions $n_{k,\ell}$ satisfy moment conditions 
\begin{subequations}
\begin{align}
\label{tree}
&\sum_{k\geq 1} kn_k =\sum_{0\le \ell<k<\infty}kn_{k,\ell}=2, \\
\label{n1-sum}
&\sum_{\ell\geq 0}\ell m_\ell=\sum_{0\le \ell<k<\infty}\ell n_{k,\ell}=n_{1}. 
\end{align}
\end{subequations}
The number of edges in a tree with $N$ vertices is $N-1$ and the total degree is twice the number of edges, $2(N-1)$, leading to \eqref{tree}. Equation \eqref{n1-sum} reflects that the mean leafdegree is the asymptotic leaf fraction $n_1:=n_{1,0}$, i.e., the limiting value of $N_{1,0}/N$, since the total leafdegree equals the number of leaves $N_1:=N_{1,0}$. In the following, we shall also use the associated generating functions $m(z)=\sum_{\ell\ge 0}z^\ell m_\ell$ and $n(y)=\sum_{k\ge 1}y^kn_k$, as well as 
\begin{equation}
\label{eq:gdef}
g(y,z) =\sum_{0\le \ell<k<\infty}y^k z^\ell n_{k,\ell},
\end{equation}
encoding the joint distribution $n_{k,\ell}$. By construction, $g(1,1)=1$, $g(1,z)=m(z)$, and $g(y,1)=n(y)$. Other consequent useful properties include $\partial_yg(1,1)=2$, $\partial_zg(1,1)=n_1$, $\partial^2_{yz}g(1,1)=\langle k\ell\rangle$, etc. We also define
\begin{equation}
n_{k}(z)=\sum_{\ell=0}^{k-1}z^\ell n_{k,\ell}, \quad m_{\ell}(y)=\sum_{k=\ell+1}^{\infty}y^kn_{k,\ell},
\end{equation}
with limiting values $n_k(1)=n_k$, $m_\ell(1)=m_\ell$ and with 
\begin{equation}
\sum_{k\ge 1}y^kn_k(z)=\sum_{\ell\ge 0} z^\ell m_\ell(y)=g(y,z).
\end{equation}
 Some quantities are expressed more conveniently by $n_{k}(y)$ or $m_\ell(z)$ than by $g(y,z)$. For instance, the mean leafdegree as a function of degree, $\bar{\ell}(k)$, is given by 
\begin{equation}
\label{mean-leaf}
 \bar{\ell}(k)=\frac{1}{n_k}\left.\frac{dn_k(z)}{dz}\right\vert_{z=1}.
\end{equation}
 
Furthermore, $g(y,0)=\sum_{k}y^kn_{k,0}$ describes the degree statistics of  protected vertices---nonleaves without any leaves as neighbors. The overall protected fraction $p=m_0-n_1$ can also be expressed as 
\begin{equation}
p=\sum_{k\ge 2}n_{k,0}=m(0)-m'(1)=g(1,0)-\partial_zg(1,1).
\end{equation}
Furthermore, the average degree of protected vertices is
\begin{equation}
\bar{k}_p=\frac{\sum_{k\ge 2}kn_{k,0}}{\sum_{k\ge 2}n_{k,0}}=\frac{\partial_yg(1,0)-n_{1}}{m_0-n_1}\,.
\end{equation}
Hence, by obtaining the multivariate generating function $g(y,z)$, numerous properties of interest can in turn be calculated by standard techniques \cite{Wilf,Flajolet}.

Our main results (Sec.~\ref{pa}) are for preferential attachment (PA) trees in the simplest setting: ordinary degree-based preferential attachment, i.e., connection probability directly proportional to degree $k$. In this model, the joint distribution satisfies the recurrence
\begin{eqnarray}
\label{eq:recursion_pa}
&(k-1)n_{k-1,\ell-1}-(2+k+\ell)n_{k,\ell} \ \ \\
&+(\ell+1)n_{k,\ell+1}+2\delta_{k,1}\delta_{\ell,0}=0,
\end{eqnarray}

with the latter Kronecker delta term reflecting the special role of the newly introduced vertices that have degree $k=1$ and leafdegree $\ell=0$. One can recast the recurrence \eqref{eq:recursion_pa} into a partial differential equation (PDE) for the generating function $g(y,z)$
\begin{equation}
\label{pa_pde}
y(1-yz)\frac{\partial g(y,z)}{\partial y}-(1-z)\frac{\partial g(y,z)}{\partial z}+2g(y,z)=2y,
\end{equation}
whose solution has integral representation
\begin{equation}
\label{pa_g}
g(y,z)=4y\int_0^1\frac{u^2}{2-y[(1-u)^2+z(1-u^2)]}du.
\end{equation}

\begin{figure}
\includegraphics[scale=0.72,trim=15 10 0 25,clip]{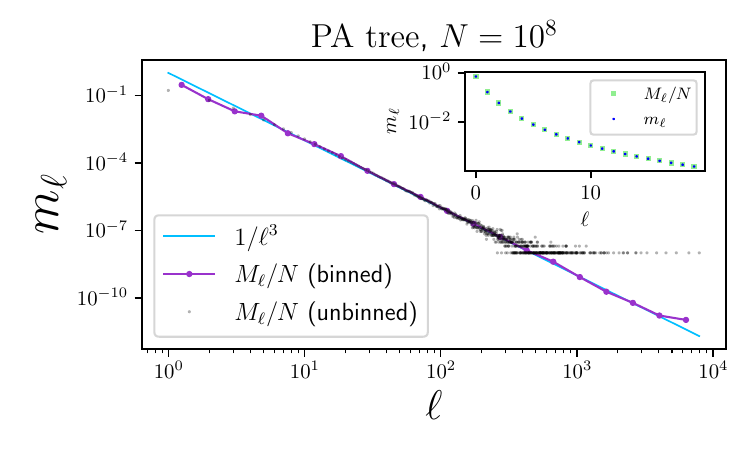}
\caption{Leafdegree distribution $M_\ell/N$ of a large PA tree ($N=10^8$), alongside theoretical values $m_\ell$ (Eq.~\eqref{eq:ml_PA}). The scattered semi-transparent dots represent the actual value of $M_{\ell}/N$, and the purple dotted line represents a log-binning thereof; the blue line is $\ell^{-3}$. Inset: comparison of $M_{\ell}/N$ (squares) and $m_{\ell}$ (dots) at small values of $\ell$.}
\label{fig:ml_pa}
\end{figure}

\begin{figure}
\ \\
\includegraphics[scale=0.72,trim=15 10 0 25,clip]{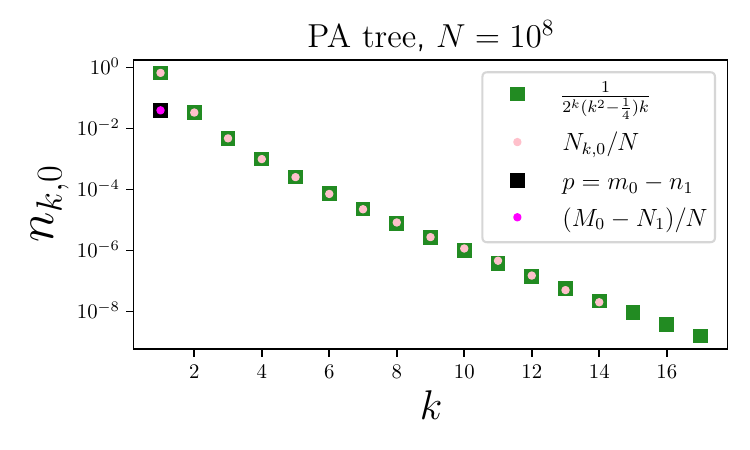}
\caption{The fraction $N_{k,0}/N$ of vertices with degree $k$ and leafdegree $0$ (pink dots) in a large PA tree ($N=10^8$), in comparison with Eq.~\eqref{eq:nk0} for $n_{k,0}$ (green squares). Also depicted at $k=1$ is the analytical value of $p$ (black square) and measured protected fraction $P/N$ (magenta dot).}
\label{fig:nk0}
\end{figure}

From this we compute numerous quantities of interest in PA trees. The leafdegree distribution is deduced from the generating function $m(z)=g(1,z)$ to yield
\begin{equation}
\label{eq:ml_PA}
m_\ell=4\int_0^1\frac{u^2}{1+2u-u^2}\left(1-\frac{2u}{1+2u-u^2}\right)^{\ell}\,du
\end{equation}
with asymptotic $m_\ell\sim 1/\ell^3$ for $\ell\gg 1$. See Fig.~\ref{fig:ml_pa}. From Eq.~\eqref{eq:ml_PA} at $\ell=0$, we obtain $m_0$, yielding the fraction $p=m_0-n_1=m_0-\frac{2}{3}$ of protected vertices
\begin{equation}
\label{eq:p_pa}
p =  \frac{12 \ \mathrm{arctanh}\left(\frac{1}{\sqrt{2}}\right)}{\sqrt{2}} -\frac{14}{3}- 4\log 2 = 0.039\, 447\, \ldots,
\end{equation}
in excellent agreement with numerical simulation of a large PA tree (Fig.~\ref{fig:nk0}). The fraction $n_{k,0}$ of vertices that have degree $k$ are protected (if $k>1$) is
\begin{equation}
\label{eq:nk0}
n_{k,0}=\frac{1}{2^k(k^2-\frac{1}{4})k}.
\end{equation}
The algebraic $k^{-3}$ contribution to the tail resembling the degree distribution $n_k=\frac{4}{k(k+1)(k+2)}$, but the dominant contribution is exponential. The exact behavior \eqref{eq:nk0} also agrees with simulation (Fig.~\ref{fig:nk0}). The average degree of a protected vertex 
\begin{equation}
\label{eq:kp_PA}
\bar{k}_{p}=\frac{\partial_yg(1,0)-\frac{2}{3}}{p}=2.202\, 493\, 667\ldots
\end{equation}
is slightly larger than the minimum degree of protected vertices, $2$. We did not find any prior works obtaining $p$ in PA, nor either $\{n_{k,0}\}_{k\ge 2}$ or $\bar{k}_p$ in any model. 
We also obtain 
\begin{equation}
\label{eq:nkz}
n_k(z)=2^{2-k}\int_0^1 t^2[(1-t)^2+z(1-t^2)]^{k-1}dt,
\end{equation}
which can be expanded in powers of $z$ to find 
\begin{equation}
\label{eq:nkl_PA}
n_{k,\ell}=\frac{\binom{k-1}{\ell}}{2^{k-2}}\int_0^1u^2(1-u)^{2(k-1)-\ell}(1+u)^\ell du,
\end{equation}
for any $k,\ell$ such that $0\le \ell<k<\infty$. The values displayed in Table~\ref{tab_nkl_pa} are extracted from \eqref{eq:nkl_PA}. Specializing \eqref{eq:nkl_PA} to the smallest leafdegree $\ell=0$ gives \eqref{eq:nk0}, while in the opposite extreme, $\ell=k-1$, we obtain
\begin{equation}
\label{nkk1}
n_{k,k-1}=\frac{(k-1)!\sqrt{\pi}}{2^k\Gamma\left(\frac{3}{2}+k\right)}\,.
\end{equation}

We also show that a typical hub, i.e., a vertex of high degree, has approximately an equal number of leaves and non-leaves among its neighbors. More precisely, we demonstrate the concentration of $\ell$ around $k/2$ as $k\rightarrow \infty$ with relative standard deviation vanishing as $1/\sqrt{k}$. This leads to the conjecture $\ell_{\mathrm{max}}/k_{\mathrm{max}}\rightarrow1/2$ in probability, with $\ell_{\mathrm{max}}$ and $k_{\mathrm{max}}$ denoting the largest values of leafdegree and degree. Indeed, in a PA tree of size $N=10^8$, we find that the highest degree vertex is also the highest leafdegree vertex, and that $\ell_{\mathrm{max}}/k_{\mathrm{max}}\approx 0.498\, 806$.

\begin{table}[h!]
\centering
\renewcommand{\arraystretch}{1.5}{
\begin{tabular}{| c | c | c | c | c | c | c | c | c |}
\hline
\diagbox{$k$}{$\ell$}  & $0$ & $1$ & $2$ & $3$ & $4$ & $5$ & $6$ & 7\\ 
\hline
$1$ & $\frac{2}{3}$  &       &     &     &   &   &  &  \\ 
\hline
$2$ & $\frac{1}{30}$   & $\frac{2}{15}$ &      &         &  &  &  &  \\
\hline
$3$ & $\frac{1}{210}$  & $\frac{1}{42}$ &  $\frac{4}{105}$ &       &  &  &  & \\
\hline
$4$ & $\frac{1}{1008}$ & $\frac{1}{168}$ & $\frac{23}{1680}$ & $\frac{4}{315}$    &  &  &  & \\
\hline
$5$ & $\frac{1}{3960}$ & $\frac{7}{3960}$ & $\frac{47}{9240}$ & $\frac{29}{3960}$ &$\frac{16}{3465}$ &   &   &  \\
\hline
$6$ & $\frac{1}{13728}$ &$\frac{1}{1716}$ & $\frac{41}{20592}$ & $\frac{19}{5148}$ &$\frac{1093}{288288}$ &$\frac{16}{9009}$  &   &  \\
\hline
$7$ & $\frac{1}{43680}$ &$\frac{3}{14560}$ & $\frac{1}{1232}$ & $\frac{29}{16016}$ &$\frac{235}{96096}$ &$\frac{309}{160160}$ & $\frac{32}{45045}$  &\\
\hline
$8$ & $\frac{1}{130560}$ &$\frac{1}{13056}$ & $\frac{193}{565760}$ & $\frac{25}{28288}$ &$\frac{139}{95744}$ &$\frac{4761}{3111680}$ & $\frac{10889}{11202048}$  & $\frac{32}{109395}$ \\
\hline
\end{tabular}
}
\caption{$n_{k,\ell}$ in the PA tree. The values $n_{k,\ell}$ are rational, which is not obvious from the integral representation \eqref{eq:nkl_PA} of the fractions, but follows from the defining recurrence Eq.~\eqref{eq:recursion_pa}. In the table, we display $n_{k,\ell}$  with $0\leq \ell <k\leq 8$.} 
\label{tab_nkl_pa}
\end{table}

To go beyond deterministic approximations, we recognize that $N_{k,\ell}$ are random variables and that these random variables are not statistically independent.  In Sec.~\ref{fluctuations}, we demonstrate the extensivity and asymptotic self-averaging of the quantities $N_{k,\ell}$ at a few values of $k,\ell$ and characterize their correlations. These features hold when the degrees $k$ and $\ell$ are kept finite while $N\to\infty$; the largest degree of the PA trees scales as $\sqrt{N}$, so extensivity of $N_k$ certainly does not hold in that range (by definition, $N_{k_{\mathrm{max}}}\approx 1$). Previous works have established concentration of degree counts $\{N_{k}\}_{k\ge 1}$ in the range $0<k<N^{1/15}$ \cite{https://doi.org/10.1002/rsa.1009}. 

In Sec.~\ref{fluctuations}, we specifically analyze $N_1, N_2$ and $N_{2,1}$. The number of leaves $N_1$ satisfies a closed stochastic equation, so one achieves a rather complete understanding of this random variable. The random variable $N_2$ is correlated with $N_1$, but the pair $(N_1,N_2)$ satisfies a closed stochastic equation. Similarly, the pair $(N_1, N_{2,1})$ satisfies a closed stochastic equation. In these situations, we establish the essential properties of joint normality and concentration that we conjecture for $N_{k,\ell}$ at all $k,\ell$. The leading behavior of all these quantities is linear in $N$, confirming extensivity and asymptotic self-averaging and also suggesting asymptotic multivariate normality.

In particular, our exact results for the averages $\langle N_1\rangle$, $\langle N_2\rangle$ and the variances $\mathbb{V}[N_1]$, $\mathbb{V}[N_2]$, suggest
\begin{subequations}
\begin{align}
\label{Gauss-1}
&P_N(N_1) \simeq \sqrt{\frac{9}{2\pi N}}\exp\!\left\{-\frac{9(N_1-\frac{2}{3}N)^2}{2N}\right\}, \\
\label{Gauss-2}
&P_N(N_2) \simeq  \sqrt{\frac{90}{23\pi N}}\exp\!\left\{-\frac{90(N_2-\frac{1}{6}N)^2}{23N}\right\}.
\end{align}
\end{subequations}
In Appendix~\ref{ap:Gauss}, we derive the Gaussian distribution for the number of leaves, Eq.~\eqref{Gauss-1}. Deriving Eq.~\eqref{Gauss-2} is more challenging since the evolution of $N_2$ depends on $N_1$. The closed stochastic equation for $(N_1, N_2)$ allows establishment of $P_N(N_1, N_2)$, the joint probability distribution of the pair $(N_1,N_2)$. We derive the covariance $\langle\!\langle N_1 N_2\rangle\!\rangle=\langle N_1 N_2\rangle - \langle N_1\rangle \langle N_2\rangle$. The probability distribution $P_N(N_1, N_2)$ is expected to be asymptotically Gaussian 
\begin{subequations}
\begin{align}
\label{Gauss-12}
P_N(N_1,N_2) \simeq \frac{1}{2\pi}\sqrt{\frac{2700}{17 N^2}}\exp\!\left\{-\frac{\Phi(N_1,N_2)}{2N}\right\}
\end{align}
with
\begin{eqnarray}
\label{eq:Phi}
\Phi(N_1,N_2) &=& \frac{345}{17}\left(N_1-\frac{2N}{3}\right)^2+ \frac{300}{17}\left(N_2-\frac{N}{6}\right)^2 \nonumber \\
&+&\frac{480}{17}\left(N_1-\frac{2N}{3}\right)\left(N_2-\frac{N}{6}\right)
\end{eqnarray}
\end{subequations}
fixed by the leading asymptotics of the averages $\langle N_1\rangle$ and $\langle N_2\rangle$, the variances $\mathbb{V}[N_1]$ and $\mathbb{V}[N_2]$, and the covariance $\langle\!\langle N_1 N_2\rangle\!\rangle$. 

The simplest nontrivial category taking into account both degree and leafdegree is the set of degree-$2$, leafdegree-$1$ vertices, the number of which is $N_{2,1}$. One such vertex arises whenever an arriving vertex attaches to a leaf. We compute the average and variance of $N_{2,1}$, suggesting an asymptotic Gaussian form
\begin{align}
P_N(N_{2,1}) \simeq 
 \sqrt{\frac{300}{49N\pi}}\exp\!\left\{-\frac{300(N_{2,1}-\frac{2}{15}N)^2}{49 N}\right\}.
\end{align}
The pair $(N_1, N_{2,1})$ satisfies a closed stochastic equation which we use to determine the covariance  $\langle\!\langle N_1 N_{2,1}\rangle\!\rangle$. The probability distribution $P_N(N_1, N_{2,1})$ is expected to be asymptotically Gaussian, of the form
\begin{equation}\begin{aligned}
P_N(N_1,N_{2,1}) \simeq \frac{45\sqrt{2}}{\pi N \sqrt{97}}\,\exp\!\left[-\frac{\Psi(N_1,N_{2,1})}{2N}\right],
\end{aligned}\end{equation}
with $\Psi(N_1,N_{2,1})$ fixed by the leading asymptotics of the averages $\langle N_1\rangle$ and $\langle N_{2,1}\rangle$, variances $\langle\!\langle N_1^2\rangle\!\rangle$ and $\langle\!\langle N_{2,1}^2\rangle\!\rangle$, and the covariance $\langle\!\langle N_1 N_{2,1}\rangle\!\rangle$.  These results extend previous works on normality and self-averaging of various quantities of interest in preferential attachment trees \cite{https://doi.org/10.1002/rsa.1009} and related sparse random graphs \cite{https://doi.org/10.1002/rsa.20905}.

In Appendix~\ref{rrt}, we further demonstrate the approach developed herein by analysis of the random recursive tree (RRT) or uniform attachment model. Both $n_k=2^{-k}$ \cite{KRB} and $m_\ell=\gamma(\ell+1,1)/\Gamma(\ell+1)$ \cite{hartle2025statistics} have been previously obtained; the joint distribution satisfies
\begin{equation}\begin{aligned}
\label{eq:recursion_rrt}
n_{k-1,\ell-1}-(2+\ell)n_{k,\ell}+(\ell+1)n_{k,\ell+1}+\delta_{k,1}\delta_{\ell,0}=0,
\end{aligned}\end{equation}
leading to PDE for $g(y,z)$, namely,
\begin{equation}\begin{aligned}
\label{rrt_pde}
(1-z)\frac{\partial g(y,z)}{\partial z}-(2-xy)g(y,z)=y.
\end{aligned}\end{equation}
The solution of \eqref{rrt_pde} admits an integral representation
\begin{equation}
\label{rrt_g}
g(y,z) =y\int_0^1 e^{-y(1-v)(1-z)}v^{1-y}dv
\end{equation}
from which we recover the known generating functions $n(y)$ and $m(z)$, the distributions $\{n_k\}_{k\ge 1}$ and $\{m_\ell\}_{\ell\ge 0}$, and the known protected fraction $p=\frac{1}{2}-\frac{1}{e} $ \cite{mahmoud2015asymptotic}. Beyond these we obtain the degree-leafdegree correlation $\langle k\ell\rangle=7/4$, the degree-stratified protected fractions $(n_{k,0})_{k\ge 2}$, protected average degree $\bar{k}_p\approx 0.266$, the conditional mean $\bar{\ell}(k)=2-3(2/3)^k$, and the convergence of degree-conditional leafdegree distribution to a Poisson of mean $2$ at large values of degree. We also obtain an exact albeit cumbersome expression for $n_{k,\ell}$ and consequent results such as $n_{k,k-1}$.

Finally, in Appendix~\ref{app:crmodel}, we provide analysis of a model recently introduced in Ref.~\cite{QPA}, where a two-index recurrence was derived for a joint distribution equivalent to $n_{k,\ell}$. In terms of $n_{k,\ell}$, that recurrence becomes
\begin{eqnarray}
\label{rec:LPA-0-maintext}
(2\ell+1)n_{k,\ell} &=& (\ell-1)n_{k-1,\ell-1}+(\ell+1)n_{k,\ell+1} \nonumber \\
&+&\tfrac{1}{2}(\delta_{k,1}\delta_{\ell,0} + \delta_{k,2}\delta_{\ell,1}).
\end{eqnarray}
Although this model was obtained in a limiting parameter regime of a two-parameter model, we note that it has a natural and self-contained definition. Starting from a tree of size $3$, iterate the following attachment rule for each arriving vertex $j=4,\ldots,N$:
\begin{itemize}
    \item Select a leaf $i$ uniformly at random.
    \item Attach to $i$ with probability $\frac{1}{2}$; else to its neighbor.
\end{itemize}
By this mechanism, the attachment probability to nonleaves is proportional to leafdegree, and thus protected vertices are never attached to. A leaf-based description naturally applies; summing Eq.~\eqref{rec:LPA-0-maintext} over $k>\ell$ yields a closed recurrence for the leafdegree distribution:
\begin{equation}
\begin{aligned}
\label{ml_CR}
    (2\ell+1)m_\ell=(\ell-1)m_{\ell-1}+(\ell+1)m_{\ell+1}+\frac{\delta_{\ell,0}+\delta_{\ell,1}}{2},
\end{aligned}
\end{equation}
which can also be deduced directly without appeal to $n_{k,\ell}$, as we demonstrate in Sec.~\ref{app:crmodel_recu}. From Eq.~\eqref{ml_CR} we obtain the leafdegree distribution generating function $m(z)$ (Sec.~\ref{app:crmodel_gf}), and by extracting coefficients, we obtain $m_\ell$ for arbitrary $\ell=0,1,2,\ldots$; notable results include the strange equality of $p$ and $m_1$ and their value:
\begin{equation}
p=m_1=\frac{1+e\, \mathrm{Ei}(-1)}{2}=0.201\, 826\, \ldots,
\end{equation}
with $\mathrm{Ei}(x)=\int_{-\infty}^xt^{-1}e^tdt$ the exponential integral; also, that the leafdegree distribution has stretched exponential tail:
\begin{equation}
    m_\ell\sim \ell^{-1}e^{-2\sqrt{\ell}}.
\end{equation}
Finding $n_{k,\ell}$, and consequently $n_k$ and $n_{k,0}$, are challenging tasks. In Appendix~\ref{app:crmodel_joint} we derive an (unwieldy) expression for $g(y,z)$ and consequently $g(y,1)$ and $g(y,0)$, but we do not analytically determine their expansion coefficients. We nevertheless examine $n_{k}$ by exact numerical solution of the two-index recurrence Eq.~\eqref{rec:LPA-0-maintext} and validate the results in simulation.

\section{Preferential attachment}
\label{pa}

In this section, we derive for PA the probability generating functions $g(y,z)$, $m(z)$, $m_\ell(y)$, and $n_k(z)$, and recover the known $n(y)$. We obtain explicit formulae for the joint distribution $n_{k,\ell}$ and for the leafdegree distribution $m_\ell$. We also obtain the protected fraction $p$, the protected degree fractions $(n_{k,0})_{k\ge 2}$, and the conditional mean $\bar{\ell}(k)$, and demonstrate the associated concentration. We begin with a derivation of the recursion for $n_{k,\ell}$ in PA and the resulting PDE for $g(y,z)$ and its solution.

\subsection{Multivariate generating function}

From consideration of the PA process, the probability of attaching to a vertex of degree $k$ and leafdegree $\ell$ is $kN_{k,\ell}/\sum_{k',\ell'}k'N_{k',\ell'}=kN_{k,\ell}/2(N-1)$. Direct attachments lead to $k\rightarrow k+1$ and $\ell\rightarrow\ell+1$, whereas attachment to a neighbor that is a leaf leads to $\ell\rightarrow\ell-1$ (but does not affect degree; $k\rightarrow k$). Therefore,
\begin{eqnarray}
\frac{dN_{k,\ell}}{dN} &=&\frac{(k-1)N_{k-1,\ell-1}}{2(N-1)}+\frac{(\ell+1)N_{k,\ell+1}}{2(N-1)}-\frac{kN_{k,\ell}}{2(N-1)}  \nonumber \\
&-&\frac{\ell N_{k,\ell}}{2(N-1)}+\delta_{k,1}\delta_{\ell,0}.
\end{eqnarray}
With extensive scaling and asymptotic self-averaging, $dN_{k,\ell}/dN$ approaches the intensive limiting fraction $n_{k,\ell}$. Hence, we arrive at the recurrence
\begin{eqnarray}
n_{k,\ell}&=&\frac{(k-1)n_{k-1,\ell-1}}{2}+\frac{(\ell+1)n_{k,\ell+1}}{2}-\frac{kn_{k,\ell}}{2} \nonumber \\
&-&\frac{\ell n_{k,\ell}}{2}+\delta_{k,1}\delta_{\ell,0},
\end{eqnarray}
which reduces to Eq.~\eqref{eq:recursion_pa}. The boundary conditions are $n_{k,-1}=0$, $n_{\infty,\ell}=0$, and $n_{q,q}=0$. The non-zero fractions, $n_{k,\ell}>0$ are thus those with $0\le \ell<k<\infty$.

The generating function $g(y,z)=\sum_{0\le \ell<k<\infty}n_{k,\ell}y^kz^\ell$ transforms the two-index recursion for $n_{k,\ell}$ into a  partial differential equation (PDE) for $g(y,z)$. We multiply Eq.~\eqref{eq:recursion_pa} by $z^\ell y^k$, sum over all $0\le \ell<k<\infty$, and arrive at the announced Eq.~\eqref{pa_pde}. In the derivation, we have used identities following from the definition of $g$:
\begin{equation}\begin{aligned}
 \sum_{k >\ell\geq 0} kn_{k,\ell}\,y^k z^\ell &
 =y \partial_y g\\
 \sum_{k >\ell\geq 0} \ell n_{k,\ell}\,y^k z^\ell &=
 z \partial_z g,\\ 
 \sum_{k >\ell\geq 0} (k-1)n_{k-1,\ell-1}\,y^{k} z^\ell &
 =y^2z \partial_y g,\\
 \sum_{k >\ell\geq 0} (\ell+1)n_{k,\ell+1}\,y^k z^\ell &=
 \partial_z g, 
\end{aligned}\end{equation}
with $\partial_z:=\frac{\partial}{\partial z}$ and $\partial_y:=\frac{\partial}{\partial y}$. We apply the method of characteristics to solve the PDE for $g(y,z)$. We first rewrite the PDE as
 \begin{equation}\begin{aligned}
y(1-yz)\partial_yg-(1-z)\partial_zg+2g=2y,
\end{aligned}\end{equation}
Characteristics $y(z)$ solve
\begin{equation}\begin{aligned}
\frac{dy}{dz}=-\frac{y(1-yz)}{1-z},
\end{aligned}\end{equation}
from which
\begin{equation}\begin{aligned}
y(z)=-\frac{2 (1 - z)}{z^2 + 2 C}.
\end{aligned}\end{equation}
Along characteristics,
\begin{equation}\begin{aligned}
\frac{dg}{dz}-\frac{2g}{1-z}=\frac{4}{z^2 + 2 C}.
\end{aligned}\end{equation}
Integrating factor $(1-z)^2$ yields 
\begin{equation}\begin{aligned}
\frac{d}{dz}\left[(1-z)^2g(z)\right]=\frac{4(1-z)^2}{z^2 + 2 C},
\end{aligned}\end{equation}
from which the integral solution can be written
\begin{equation}\begin{aligned}
g(y,z)=\frac{1}{(1-z)^2}\int_1^z\frac{4(1-t)^2}{t^2 +2C}\,dt,
\end{aligned}\end{equation}
where $C=-z^2/2- (1 - z)/y$. Thus
\begin{equation}\begin{aligned}
\label{gyz:sol}
g(y,z)=\frac{4y}{(1-z)^2}\int_1^z\frac{(1-t)^2}{y(t^2 -z^2)-2(1-z)}\,dt.
\end{aligned}\end{equation}
This generating function is the source of many subsequent results. 

In some calculations, it proves convenient to re-express \eqref{gyz:sol} in a different form using a change of variables. Letting $u=(1-t)/(1-z)$, we obtain
\begin{equation}\begin{aligned}
\label{gyz-u}
g(y,z)=4y\int_0^1\frac{u^2}{2-y[(1-u)^2+z(1-u^2)]}\,du.\\
\end{aligned}\end{equation}
One may verify that $g(y,z)$ has the requisite properties encoding normalization and the moment constraints:
\begin{equation}\begin{aligned}
g(1,1)&=1,\\
\partial_y g(1,1)&=2,\\ 
\partial_zg(1,1)&=2/3.
\end{aligned}\end{equation}
Expansion in powers of $y$ to obtain integral formulae for $n_k(z)$ is straightforward, as is subsequent power series in $z$ to obtain expressions for $n_{k,\ell}$; see Sec.~\ref{ssec:nkz_mky} and ~\ref{ssec:nkl}.

\subsection{Degree and leafdegree generating functions}
\label{ssec:ml_nk_pa}

We first consider the leaf degree distribution. Specializing \eqref{gyz-u} to $y=1$ gives the generating function $m(z)=g(1,z)$:
\begin{equation}\begin{aligned}
\label{mz-u}
m(z)=4\int_0^1\frac{u^2}{2-(1-u)^2-z(1-u^2)}\,du.\\
\end{aligned}\end{equation}
One can explicitly evaluate the integral in terms of arctan, log, and powers of $1-z$, but the integral form of $m(z)$ is more convenient. Expanding in powers of $z$ yields $m_\ell$ as the coefficient of $z^\ell$. Rewriting \eqref{mz-u} as
\begin{equation}\begin{aligned}
\label{mz-u2}
m(z)&=4\int_0^1\frac{u^2}{1+2u-u^2}\frac{1}{1-\frac{z(1-u^2)}{1+2u-u^2}}\,du\\
\end{aligned}\end{equation}
and expanding in powers of $z$ gives
\begin{equation}\begin{aligned}
m_\ell=4\int_0^1\frac{u^2}{1+2u-u^2}\left(\frac{1-u^2}{1+2u-u^2}\right)^\ell\, du,
\end{aligned}\end{equation}
equivalent to Eq.~\eqref{eq:ml_PA}. The fraction $m_0=m(0)=g(1,0)$ is thus 
\begin{eqnarray}
m_0&=&4\int_0^1\frac{u^2}{1+2u-2u^2}\,du\nonumber \\
&=& \frac{12 \ \mathrm{arctanh}\left(\frac{1}{\sqrt{2}}\right)}{\sqrt{2}} - 4(1+\log 2),
\end{eqnarray}
leading to Eq.~\eqref{eq:p_pa}. At $\ell=1,2$, we have
\begin{eqnarray*}
m_1&=&11 \sqrt{2} \mathrm{arctanh}\left(\frac{1}{\sqrt{2}}\right)- 8 (1+\log 2) =0.165\, 778\, \ldots\\
m_2&=&\frac{67\sqrt{2}}{4}  \mathrm{arctanh}\left(\frac{1}{\sqrt{2}}\right) -\frac{25}{2} - 12 \log 2=0.060\, 279\, \ldots
\end{eqnarray*}
and so on. Numerical values of $m_\ell$ at some higher values of $\ell$ are
\begin{equation}\begin{aligned}
m_3&=0.027\, 253 \, \ldots\\
m_4&=0.014\, 147\, \ldots \\
m_5&=0.008\, 098\, \ldots\\
m_6&=0.004\, 989\, \ldots\\
m_7&=0.003\, 255\, \ldots\\
m_8&=0.002\, 224\, \ldots\\
m_9&=0.001\, 578\, \ldots\\
m_{10}&=0.001\, 156\, \ldots\\
\end{aligned}\end{equation}
which align well with simulation data even for a single large graph (Fig.~\ref{fig:ml_pa}).
Setting $z=1$ yields
\begin{equation}\begin{aligned}
n(y)=g(y,1)=2y\int_0^1\frac{u^2}{1-y(1-u)}du.
\end{aligned}\end{equation}
Evaluation of the integral yields 
\begin{equation}\begin{aligned}
n(y)=3-\frac{2}{y}-2\left(1-\frac{1}{y}\right)^2\log(1-y).
\end{aligned}\end{equation}
Expanding in powers of $y$, the $k$th coefficient is
\begin{equation}\begin{aligned}
n_k=\frac{4}{k(k+1)(k+2)},
\end{aligned}\end{equation}
which is the well-known solution \cite{Simon55,KR00,Sergey00}.

\subsection{Protected vertices in preferential attachment}

For the protected degree fractions $\{n_{k,0}\}_{k\ge 1}$, the corresponding generating function $g(y,0)=\sum_{k\ge 1}n_{k,0}y^k$. Using \eqref{gyz:sol} we find
\begin{equation}\begin{aligned}
g(y,0)=2y\int_0^1\frac{t^2}{1-(y/2)(1-t)^2}\,dt,
\end{aligned}\end{equation}
which is rearranged as
\begin{equation}\begin{aligned}
g(y,0)=4\int_0^1\frac{t^2}{(1-t)^2}\frac{(y/2)(1-t)^2}{1-(y/2)(1-t)^2}\,dt.\\
\end{aligned}\end{equation}
Expanding this generating function in powers of $y$ and using $\int_0^1t^2(1-t)^{2k-2}\,dt=1/(4k^3-k)$, we arrive at the announced result \eqref{eq:nk0}.

The average degree of protected vertices in PA is 
\begin{equation}\begin{aligned}
\bar{k}_p=\frac{\sum_{k\geq 2} k n_{k,0}-\frac{2}{3}}{p}=2.202\, 493\, 667\, \ldots,
\end{aligned}\end{equation}
where $\sum_{k\geq 2} k n_{k,0}=\frac{4}{3}-\sqrt{2} \text{arcsinh}(1)$ is calculable directly using Eq.~\eqref{eq:nk0}.

\subsection{Generating functions $n_k(z)$ and $m_\ell(y)$}
\label{ssec:nkz_mky}

Expanding $g(y,z)$ in powers $y^k$ ($k\ge 1$) or in powers $z^\ell$ ($\ell\ge 0$) gives $n_k(z)$ or $m_\ell(y)$, respectively. To obtain $m_\ell(y)$, we first rewrite \eqref{gyz:sol} as
\begin{equation*}
g(y,z)=4y\int_0^1\frac{t^2}{2-(1-t)^2y}\frac{1}{1-\frac{(1-t^2)zy}{2-(1-t)^2y}}\,dt,
\end{equation*}
and then use the geometric series to obtain
\begin{equation}
m_\ell(y)=4y^{1+\ell}\int_0^1\frac{t^2}{2-(1-t)^2y}\left(\frac{(1-t^2)}{2-(1-t)^2y}\right)^\ell\, dt.
\end{equation}
From $m_\ell(y)$, one gets $m_\ell=\lim_{y\rightarrow 1}m_\ell(y)$; one can also obtain $n_{k,\ell}$ by expanding in powers of $y$. Similarly, we rewrite \eqref{gyz:sol} as
\begin{equation}\begin{aligned}
g(y,z)&=2y\int_0^1\frac{t^2}{1-[(1-t)^2+z(1-t^2)](y/2)}\,dt\\
&=\frac{y^{q+1}}{2^{q-1}}\int_0^1t^2\sum_{q=0}^\infty [(1-t)^2+z(1-t^2)]^q\,dt,
\end{aligned}\end{equation}
from which we extract 
\begin{equation}
n_k(z)=2^{2-k}\int_0^1t^2[(1-t)^2+z(1-t^2)]^{k-1}\,dt.
\end{equation}
The values of $n_{k,\ell}$ can be obtained by expansion of $n_{k}(z)$ in powers of $z$; see Sec.~\ref{ssec:nkl}.

\subsection{Mean and variance of the leafdegree of vertices with fixed degree}

Here we derive the mean and variance of the leafdegree of vertices with fixed degree $k$. We also explore the concentration of the leaf degree for vertices with $k\gg 1$. (Similar concentration phenomenon was conjectured in Ref.~\cite{hartle2025growing}, but in a leaf-proliferating regime where $\ell/k\rightarrow 1$ for $k\gg 1$.) 

By definition
\begin{equation}
\label{mean-leaf-2}
\bar{\ell}(k)=
\frac{\sum_{\ell=0}^{k-1}\ell n_{k,\ell}}{n_k}=\left.\frac{n'_k(z)}{n_k(z)}\right\vert_{z=1}
\end{equation}
which evaluates to
\begin{equation}
\label{l-av:k}
\bar{\ell}(k)=1+\frac{k}{2}-\frac{6}{k+3}.
\end{equation}
The average leafdegree ratio for $k\gg 1$ thus behaves as 
\begin{equation}\begin{aligned}
\frac{\bar{\ell}(k)}{k}=\frac{1}{2}+O(1/k).
\end{aligned}\end{equation}
Secondly, we evaluate
\begin{equation}\begin{aligned}
\langle \ell(\ell-1)\rangle_k&:=\sum_{\ell\ge 0}\ell(\ell-1)\frac{n_{k,\ell}}{n_k}=\frac{1}{n_k}\left.\frac{d^2n_k(z)}{dz^2}\right\vert_{z=1}\\
&=\frac{(k-2) (k-1) (48 + 13 k + k^2)}{
 4 (k+3) (k+4)}.
\end{aligned}\end{equation}
The variance of $\ell$ at a fixed $k$ reads
\begin{equation}\begin{aligned}
\mathbb{V}[\ell|k]&=\langle \ell^2\rangle_k-\bar{\ell}(k)^2=\langle \ell(\ell-1)\rangle_k+\bar{\ell}(k)-\bar{\ell}(k)^2\\
&=\frac{(k-1) k (18 + 13 k + k^2)}{4 ( k+3)^2 (k+4)}.
\end{aligned}\end{equation}
Expanding at large $k$ gives 
\begin{equation}
\mathbb{V}[\ell|k]=\frac{k}{4}+\frac{1}{2}-\frac{12}{k}+O(1/k^2). 
\end{equation}
Therefore the leafdegrees of vertices with degree $k\gg 1$ concentrate around $k/2$. We anticipate normal fluctuations akin to the characterizations in Sec.~\ref{fluctuations}. More precisely, the probability distribution $\Pi(\ell|k):=n_{k,\ell}/n_k$ is expected to converge to the Gaussian form 
\begin{align}
\label{Gauss-leaf}
\Pi(\ell|k) \simeq \sqrt{\frac{2}{\pi k}}\exp\!\left\{-\frac{2(\ell-\frac{1}{2}k)^2}{k}\right\}
\end{align}
in the $k\to\infty$ limit.

\subsection{Fractions $n_{k,\ell}$ from $n_k(z)$}
\label{ssec:nkl}

We outlined (Sec.~\ref{ssec:nkz_mky}) various ways to deduce explicit formulae for $n_{k,\ell}$ from $g(y,z)$. One such procedure is based on expanding $n_k(z)$ in powers of $z$. We write
\begin{eqnarray*}
n_k(z)&=&2^{2-k}\int_0^1u^2[(1-u)^2+z(1-u^2)]^{k-1}du\\
&=&\sum_{\ell=0}^{k-1}\frac{\binom{k-1}{\ell}}{2^{k-2}}\,z^\ell\int_0^1u^2(1-u)^{2(k-1-\ell)}(1-u^2)^\ell du
\end{eqnarray*}
from which we deduced the announced formula \eqref{eq:nkl_PA}. Specializing \eqref{eq:nkl_PA} to $\ell=0$ and $\ell=k-1$ gives $n_{k,\ell}$ in the extreme cases, Eqs.~\eqref{eq:nk0} and \eqref{nkk1}. As a consequence of \eqref{nkk1} we find the total fraction of vertices that are near-stars:
\begin{equation}
\sum_{k\ge 1}n_{k,k-1}=4-\pi=0.858\, 407\, 346\, \ldots
\end{equation}

Similarly, specializing \eqref{eq:nkl_PA} to $\ell=1$ gives 
\begin{equation}\begin{aligned}
\label{eq:nk1}
n_{k,1}=\frac{k+2}{2^k(k^2-\frac{1}{4})k}\,.
\end{aligned}\end{equation}
The dual formula
\begin{equation}\begin{aligned}
\label{nkk2}
n_{k,k-2}=\frac{(k-1)!}{2^k}\left[\frac{\sqrt{\pi}}{\Gamma\left(\frac{1}{2}+k\right)}+ \frac{3\sqrt{\pi}}{2\Gamma\left(\frac{3}{2}+k\right)}-\frac{4}{k!}\right]
\end{aligned}\end{equation}
follows from \eqref{eq:nkl_PA} at $\ell=k-2$.  Both \eqref{eq:nk1} and \eqref{nkk2} are valid for all $k\geq 2$. In Appendix~\ref{rrt} we provide similar analyses for the RRT.

\section{Asymptotic Self-averaging in preferential attachment}
\label{fluctuations}

The basis of this work is extensivity and self-averaging of the variables $\{N_{k,\ell}\}$ in random PA trees $G_N$ as $N\rightarrow\infty$. The convergence in probability of $N_{k,\ell}/N$ to a nonrandom limiting values allows construction of recurrences among $n_{k,\ell}$. In this section, we demonstrate this convergence properties in a variety of cases. The average $\langle N_{k,\ell}\rangle$ and the variance $\mathbb{V}[N_{k,\ell}]=\langle N_{k,\ell}^2\rangle-\langle N_{k,\ell}\rangle^2$ both scale linearly with $N$, when $k$ and $\ell$ are kept finite:
\begin{equation}
\label{av-var}
\lim_{N\to\infty}N^{-1}
\begin{bmatrix}
\langle N_{k,\ell}\rangle  \\
\mathbb{V}[N_{k,\ell}]
\end{bmatrix}
= \begin{bmatrix}
n_{k,\ell}\\
\chi_{k,\ell}
\end{bmatrix}.
\end{equation}
Because of the Markovian nature of the PA process, we anticipate that $P_N(N_{k,\ell})$ approaches, in the $N\to\infty$ limit, a Gaussian distribution for every fixed $k$ and $\ell$:
\begin{equation}
\label{Gauss}
P_N(N_{k,\ell}) \simeq \frac{1}{\sqrt{2\pi N \chi_{k,\ell}}}\exp\!\left\{-\frac{(N_{k,\ell}-N n_{k,\ell})^2}{2N\chi_{k,\ell}}\right\}.
\end{equation}
The average and the variance are the first two cumulants. Higher cumulants are also expected to scale linearly with $N$. For instance 
\begin{equation}
\label{cum-p}
\lim_{N\to\infty}N^{-1}\langle\!\langle N_k^p\rangle\!\rangle = \kappa^{(p)}_k.
\end{equation}
when $k$ is kept finite. Recalling the definition of the cumulants we can rewrite \eqref{cum-p} as
\begin{equation}
\label{cum-p:def}
\lim_{N\to\infty}N^{-1} \log \left\langle e^{xN_k(N)}\right \rangle = \sum_{p\geq 1}\kappa^{(p)}_k\,\frac{x^p}{p!}.
\end{equation}
Random quantities $N_i$ and $N_j$ are correlated. Correlations are captured by the two-variate cumulant generating function, which is also expected to exhibit a linear scaling with $N$:
\begin{equation}
\label{cum-pq:def}
\lim_{N\to\infty}N^{-1} \log \left\langle e^{xN_i+yN_j}\right \rangle = \sum_{p,q\geq 1}\kappa^{(p,q)}_{i j}\,\frac{x^p\,y^q}{p!\,q!}.
\end{equation}

We do not establish these results in full generality. We limit ourselves to $k=1$ and $k=2$. However, the analysis can be extended to higher $k$. Focusing on the pure degree distribution, we confirm that the averages $\langle N_k\rangle$ and covariances $\langle\!\langle N_i N_j\rangle\!\rangle$ scale linearly with $N$
\begin{equation}
\label{av-covar}
\lim_{N\to\infty}N^{-1}
\begin{bmatrix}
\langle N_k\rangle  \\
\langle\!\langle N_i N_j\rangle\!\rangle
\end{bmatrix}
= \begin{bmatrix}
n_k\\
\sigma_{ij}
\end{bmatrix}.
\end{equation}
for $k\leq 2$ and $i,j\leq 2$. The covariances are defined by $\langle\!\langle N_i N_j\rangle\!\rangle = \langle N_i N_j\rangle-\langle N_i\rangle\langle N_j\rangle$, and the normalized covariances can be alternatively written as $\sigma_{ij}=\kappa^{(1,1)}_{i j}$. We determine normalized covariances
\begin{equation}
\sigma_{11} = \frac{1}{9}\,, \quad \sigma_{22} = \frac{23}{180}\,, \quad \sigma_{12} = \sigma_{21} = -\frac{4}{45}.
\end{equation}

We confirm the linear growth with $N$ of the cumulants of the total number of leaves $N_1$. In this case, we determine the cumulant generating function. 

Similarly, we provide analysis of the joint distribution of $(N_1,N_{2,1})$. The variance parameter for $N_1$ remains the same, and the covariance $\sigma_{1,(2,1)}=\sigma_{(2,1),1}$ of $(N_1,N_{2,1})$ and variance $\sigma_{(2,1),(2,1)}$ of $N_{2,1}$ are determined to be
\begin{equation}\begin{aligned}
\sigma_{1,(2,1)}=\sigma_{(2,1),1}=-\frac{1}{18}, \ \ \sigma_{(2,1),(2,1)}=\frac{49}{600}.
\end{aligned}\end{equation}

Some results of Sec.~\ref{subsec:leaves} and Sec.~\ref{subsec:2} can be extracted from Ref.~\cite{KR02-fluct}, where fluctuations in the same PA model, but with a different initial condition, have been studied.

\subsection{Number of leaves}
\label{subsec:leaves}

The total number of leaves $N_{1,0}=N_1$ varies in a PA tree according to stochastic equation
\begin{equation}
\label{N1N}
N_1(N+1)=
\begin{cases}
N_1      & \text{prob}~~\frac{N_1}{2(N-1)}, \\
N_1 +1 & \text{prob}~~1-\frac{N_1}{2(N-1)},
\end{cases}
\end{equation}
for $N\geq 2$. (We write $N_1$ instead of $N_1(N)$ when it would not lead to confusion.)

The total number of leaves is deterministic quantity when $N=2$ and $N=3$, viz., $N_1(2)=N_1(3)=2$, and random for $N\geq 4$. For instance,
\begin{align}
\label{L4}
N_1(4) =
\begin{cases}
2 & \text{prob}~~\frac{1}{2} \\
3 & \text{prob}~~\frac{1}{2}
\end{cases}.
\end{align}

Averaging \eqref{N1N} we find
\begin{equation}
\label{N1:av}
\langle N_1(N+1) \rangle=\left[1-\frac{1}{2(N-1)}\right]\langle N_1\rangle + 1.
\end{equation}
Solving this recurrence subject to the initial condition $\langle N_1(3) \rangle=2$ yields
\begin{equation}
\label{eq:N1}
\langle N_1\rangle=\frac{2}{3}\,(N-1)+\frac{4}{3\sqrt{\pi}}\,\frac{\Gamma(N-\frac{3}{2})}{\Gamma(N-1)}.
\end{equation}
Similarly, one finds a recurrence for the second moment,
\begin{eqnarray}
\langle N_1^2(N+1) \rangle &= \left[1-\frac{1}{N-1}\right]\langle N^2_1\rangle \nonumber \\
&+ \left[2-\frac{1}{2(N-1)}\right]\langle N_1\rangle+ 1,
\end{eqnarray}
which is solved subject to $\langle N_1^2(3) \rangle=4$ to yield
\begin{equation}
\begin{aligned}
\label{N1N1:sol}
\langle N^2_1\rangle &=\frac{(4N-3)(N-1)}{9}\\ & +\frac{16}{9\sqrt{\pi}}\,\frac{\Gamma(N-\frac{1}{2})}{\Gamma(N-1)}-\frac{4}{3\sqrt{\pi}}\,\frac{\Gamma(N-\frac{3}{2})}{\Gamma(N-1)}.
\end{aligned}
\end{equation}
Combining \eqref{eq:N1}  and \eqref{N1N1:sol}, we find the variance
\begin{equation}
\label{N1:var}
\mathbb{V}[N_1]  = \frac{N-1}{9} -\frac{20}{9\sqrt{\pi}}\,\frac{\Gamma(N-\frac{3}{2})}{\Gamma(N-1)} + \frac{16}{9 \pi}\left[\frac{\Gamma(N-\frac{3}{2})}{\Gamma(N-1)}\right]^2. 
\end{equation}
From \eqref{eq:N1} we recover $n_1=\frac{2}{3}$, while \eqref{N1:var} gives $\sigma_{11}=\frac{1}{9}$. 

The number of leaves evolves according to {\em closed} stochastic recurrence \eqref{N1N}, so it could be feasible to rigorously derive the Gaussian distribution, and perhaps also establish the linear growth of all cumulants of $N_1$ and compute all $\kappa^{(p)}$. The stochastic equation \eqref{N1N} leads to the recurrence 
\begin{equation}\begin{aligned}
\label{PN1N}
P_{N+2}(N_1) &=\left(1-\frac{N_1-1}{2N}\right)P_{N+1}(N_1-1)\\&+ \frac{N_1}{2N}\,P_{N+1}(N_1)
\end{aligned}\end{equation}
for the probability distribution $P_N(N_1)$. In Appendix~\ref{ap:Gauss}, we show that the solution approaches to the Gaussian form \eqref{Gauss-1} when $N\gg 1$. 

Finding an exact solution of \eqref{PN1N} valid for all $N\geq 3$ is a challenge. We have successfully identified the leading behavior of the cumulant generating function. (This result is significantly stronger than proving that the distribution is asymptotically Gaussian.) In Appendix~\ref{ap:cum}, we show that
\begin{equation}
\label{cum:leaves}
\lim_{N\to\infty}N^{-1} \log \left\langle e^{xN_1(N)}\right \rangle = \log\,\frac{(e^x-1)^2}{2(e^x-1-x)}.
\end{equation}
Expanding the right-hand side of Eq.~\eqref{cum:leaves} and comparing the results with the right-hand side of Eq.~\eqref{cum-p:def} we recover $\kappa_1^{(1)}=n_1=\frac{2}{3}$ and $\kappa_1^{(2)}=\sigma_{11}=\frac{1}{9}$ and determine higher cumulants. In Table \ref{Table:kappa-1}, we display the next eight normalized cumulants. 

\begin{table}[h!]
\centering
\setlength{\tabcolsep}{2.3pt}
\renewcommand{\arraystretch}{1.5}{
\begin{tabular}{| c | c | c | c | c | c | c | c | c | }
\hline
$p$    &  3      & 4      &  5     & 6      &  7     & 8     & 9     & 10  \\ 
\hline 
$\kappa_1^{(p)}$   & $-\frac{1}{135}$   &  $-\frac{2}{135}$   &  $\frac{1}{567}$   &  $\frac{68}{8505}$    &   $-\frac{1}{1215}$  &   $-\frac{32}{3645}$  
& $\frac{307}{610425}$  & $\frac{4492}{280665}$ \\ 
\hline
\end{tabular}
}
\caption{The normalized cumulants $\kappa_1^{(p)}$ for $3\leq p \leq 10$.} 
\label{Table:kappa-1}
\end{table}

\subsection{Number of degree-two vertices}
\label{subsec:2}

The total number $N_2$ of vertices of degree two varies according to stochastic equation
\begin{equation}
\label{N2N}
N_2(N+1)=
\begin{cases}
N_2-1  & \text{prob}~~\frac{N_2}{N-1}, \\
N_2+1 & \text{prob}~~\frac{N_1}{2(N-1)},\\
N_2     & \text{prob}~~1-\frac{N_1+2N_2}{2(N-1)}.
\end{cases}
\end{equation}
We must determine $n_2$ and $\sigma_{22}$ to provide a Gaussian approximation of $P(N_2;N)$. Averaging \eqref{N2N} we find
\begin{equation}
\label{N2:av}
\langle N_2(N+1) \rangle=\left[1-\frac{1}{N-1}\right]\langle N_2\rangle + \frac{\langle N_1\rangle}{2(N-1)},
\end{equation}
By inserting \eqref{eq:N1} into \eqref{N2:av} we obtain a recurrence for $\langle N_2\rangle$ which we solved subject to $\langle N_2(3)\rangle=1$ to yield, for $N\geq 3$,
\begin{equation}
\label{N2:av-sol}
\langle N_2\rangle=\frac{N-1}{6}+\frac{4}{3\sqrt{\pi}}\,\frac{\Gamma(N-\frac{3}{2})}{\Gamma(N-1)}.
\end{equation}
We thus recover the known normalized average $n_2=\frac{1}{6}$. Similarly we find a recurrence for the second moment. Note,
\begin{eqnarray}
\label{N2Nsq}
N_2(N+1)^2=N_{2}^2+
\begin{cases}
-2N_2+1  & \text{prob}~~\frac{N_2}{N-1}, \\
2N_2+1 & \text{prob}~~\frac{N_1}{2(N-1)},\\
0     & \text{prob}~~1-\frac{N_1+2N_2}{2(N-1)}.
\end{cases}
\end{eqnarray}

Thus
\begin{equation}\begin{aligned}
\label{N22:rec}
\langle N_2^2(N+1) \rangle
&= \left[1-\frac{2}{N-1}\right]\langle N^2_2\rangle +  \frac{\langle N_1N_2\rangle}{N-1}\\
&+ \frac{\langle N_1\rangle + 2\langle N_2\rangle}{2(N-1)}.
\end{aligned}\end{equation}
The recurrence \eqref{N22:rec} is not closed---it contains yet unknown correlator $\langle N_1N_2\rangle$. To derive a governing equation for this correlator we combine \eqref{N1N} and \eqref{N2N} and find that $N_1N_2$ must always change, with $N_{1}(N+1)N_2(N+1)$ equalling
\begin{equation}
\begin{cases}
(N_1+1)(N_2-1) & \text{prob}~~ \frac{N_2}{N-1}, \\
N_1(N_2+1)       & \text{prob}~~ \frac{N_1}{2(N-1)},\\
(N_1+1)N_2       & \text{prob}~~ 1-\frac{N_1+2N_2}{2(N-1)},
\end{cases}
\end{equation}
which we average to obtain the recurrence for $\langle N_1N_2\rangle$ valid for $N\geq 3$:
\begin{equation}
\begin{aligned}
\label{N12:rec}
\langle N_1N_2\rangle_{N+1}&= \left[1-\frac{3}{2(N-1)}\right]\langle N_1N_2\rangle \\
&+ \left[1-\frac{1}{N-1}\right]\langle N_2\rangle + \frac{\langle N^2_1\rangle}{2(N-1)}.\\
\end{aligned}
\end{equation}
In contrast to \eqref{N22:rec}, the recurrence \eqref{N12:rec} is closed as we already know $\langle N_2\rangle$ and $\langle N^2_1\rangle$. Substituting \eqref{N1N1:sol} and \eqref{N2:av-sol} into \eqref{N12:rec} and solving the resulting inhomogeneous recurrence subject to $\langle N_1(3)N_2(3)\rangle=2$, we get
\begin{equation}
\begin{aligned}
\label{N12:sol}
\langle N_1 N_2\rangle &=\frac{(5N-9)(N-1)}{45}+\frac{10}{9\sqrt{\pi}}\,\frac{\Gamma(N-\frac{1}{2})}{\Gamma(N-1)}  \\
&-\frac{4}{3\sqrt{\pi}}\,\frac{\Gamma(N-\frac{3}{2})}{\Gamma(N-1)}+\frac{47}{30\sqrt{\pi}}\,\frac{\Gamma(N-\frac{5}{2})}{\Gamma(N-1)}.
\end{aligned}
\end{equation}
Using \eqref{N12:sol} together with \eqref{eq:N1} and \eqref{N2:av-sol} we compute the covariance $\langle\!\langle N_1 N_2\rangle\!\rangle = \langle N_1 N_2\rangle-\langle N_1\rangle\langle N_2\rangle$, as

\begin{equation}
\begin{aligned}
\label{cov:12}
\langle\!\langle N_1 N_2\rangle\!\rangle &=-\frac{4(N-1)}{45}-\frac{13}{9\sqrt{\pi}}\,\frac{\Gamma(N-\frac{3}{2})}{\Gamma(N-1)} \\
&+\frac{47}{30\sqrt{\pi}}\,\frac{\Gamma(N-\frac{5}{2})}{\Gamma(N-1)}- \frac{16}{9 \pi}\left[\frac{\Gamma(N-\frac{3}{2})}{\Gamma(N-1)}\right]^2.
\end{aligned}
\end{equation}
from which $\sigma_{12} =\sigma_{21} = -\frac{4}{45}$. 

Substituting \eqref{eq:N1}, \eqref{N2:av-sol} and \eqref{N12:sol} into \eqref{N22:rec} and solving the resulting inhomogeneous recurrence subject to $\langle N_2^2(3)\rangle=1$, we obtain

\begin{equation}
\begin{aligned}
\label{N22:sol}
\langle N_2^2\rangle &=\frac{(5N+18)(N-1)}{180}\\  &+\frac{2}{45\sqrt{\pi}}\,\frac{(10N^2-30N+83)\Gamma(N-\frac{5}{2})}{\Gamma(N-1)}.
\end{aligned}
\end{equation}

The variance is therefore
\begin{equation}\begin{aligned}
\label{N2:var}
\mathbb{V}[N_2] &=\frac{23(N-1)}{180} - \frac{16}{9 \pi}\left[\frac{\Gamma(N-\frac{3}{2})}{\Gamma(N-1)}\right]^2 \\ &+\frac{2}{45\sqrt{\pi}}\,\frac{(5N+58)\Gamma(N-\frac{5}{2})}{\Gamma(N-1)}, 
\end{aligned}\end{equation}
from which we deduce linear growth of variance and hence self-averaging, with variance parameter 
\begin{equation}
    \sigma_{22}=\frac{23}{180},
\end{equation}
leading to the expected asymptotic distribution \eqref{Gauss-2}. Since $N_2$ is correlated with $N_1$, the distribution $P_N(N_1,N_2)$ provides substantially more comprehensive description than $P_N(N_2)$ and $P_N(N_1)$. The expected asymptotic form of $P_N(N_1,N_2)$ is a two-variate Gaussian distribution whose form is determined by the averages and covariances. The intensive covariance matrix is
\begin{equation}
\label{Sigma}
\Sigma_2 = 
    \begin{bmatrix}
    \sigma_{11} & \sigma_{12} \\
    \sigma_{21} &  \sigma_{22}
    \end{bmatrix}
    = \renewcommand{\arraystretch}{1.5}
    \begin{bmatrix}
    \frac{1}{9}     &- \frac{4}{45} \\
    - \frac{4}{45} &  \frac{23}{180}
    \end{bmatrix},
\end{equation}
yielding the announced result \eqref{Gauss-12}.

The generalization to higher $k$ is in principle straightforward, but laborious. For instance, when $k=3$, it is again preferable to consider the triplet $(N_1,N_2,N_3)$ due to the correlations. We already know $\Sigma_2$, see \eqref{Sigma}. One thus needs to compute $\sigma_{13}, \sigma_{23},\sigma_{33}$ to obtain an explicit asymptotic form of $P_N(N_1,N_2,N_3)$. This is anticipated also because of analogous results in the RRT: there, multivariate normality of $\{N_k\}_{k\ge 0}$ has been established, including a full characterization of covariances \cite{Janson05}.

In Table \ref{Table:kappa-2}, we display the first five normalized cumulants of $N_2$. The computation of these cumulants  is relegated to Appendix~\ref{ap:cum}.

\begin{table}[h!]
\centering
\setlength{\tabcolsep}{5pt}
\renewcommand{\arraystretch}{1.5}{
\begin{tabular}{| c | c | c | c | c | c | c | c | c |}
\hline
$q$                         & 1                    & 2                           &  3                          & 4                                     &  5                       & 6  \\ 
\hline 
$\kappa_2^{(q)}$   & $\frac{1}{6}$   &  $\frac{23}{180}$ &  $\frac{13}{189}$   &$-\frac{251}{37800}$  & $-\frac{445}{6237}$  & $-\frac{9790117}{170270100}$\\ 
\hline
\end{tabular}
}
\caption{The normalized cumulants $\kappa_2^{(q)}$ for $q \leq 6$.} 
\label{Table:kappa-2}
\end{table}

\subsection{Number of post-leaf vertices $N_{2,1}$}
\label{subsec:21}

\begin{figure*}
    \centering
    \includegraphics[scale=0.25,trim=100 60 20 80,clip]{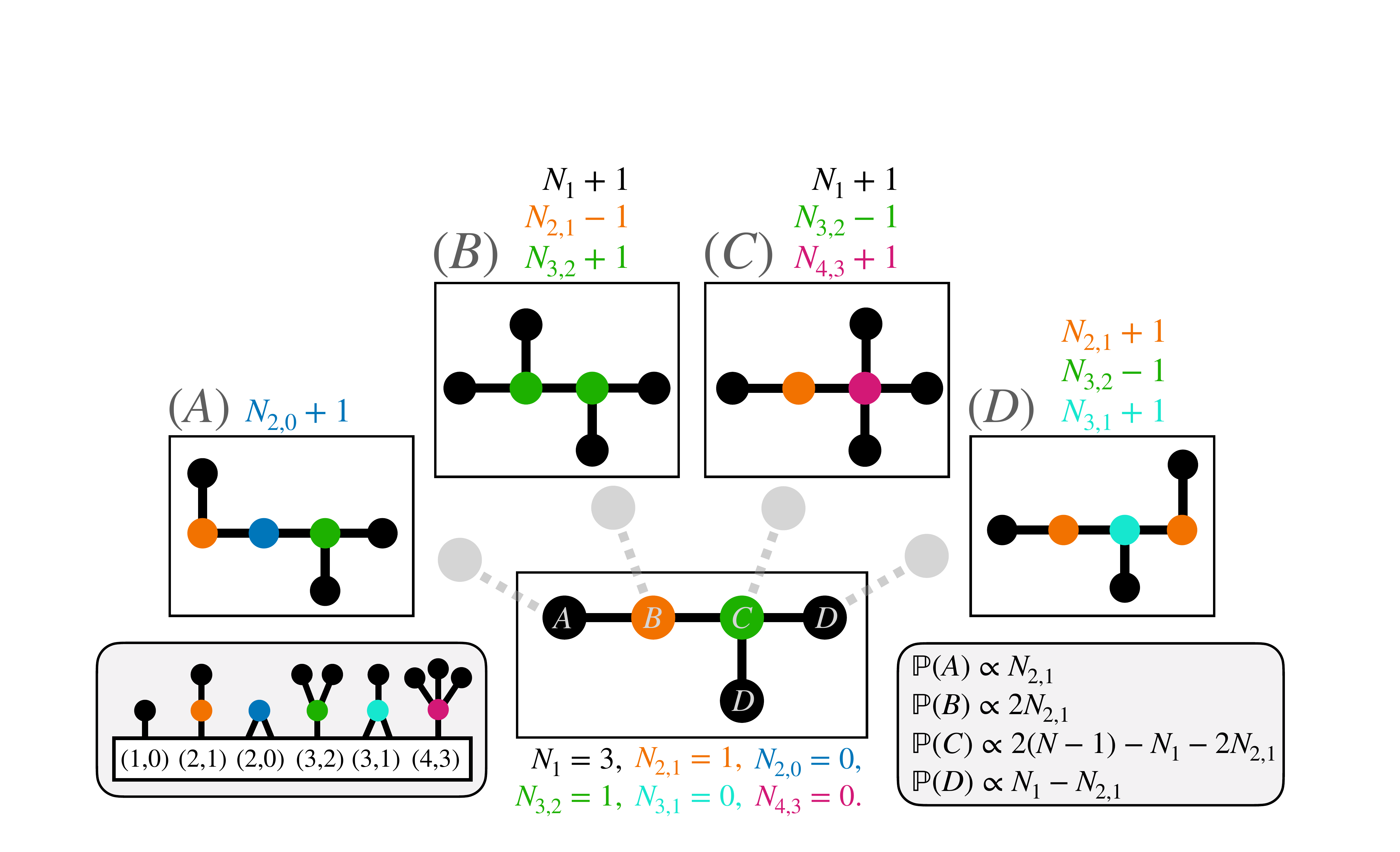}
    \caption{The four types of transitions relevant to $N_{2,1}$, exemplified by the four structurally distinct attachment options labeled A, B, C, and D, in the $5$-vertex tree depicted. Gray vertices with dashed-line edges represent the distinct attachment possibilities. (Attachment to either of the two vertices labeled $D$ yields the same structural outcome.) The panels labeled (A), (B), (C), and (D) display the resulting tree for each possible attachment. From left to right: (A) depicts attachment to the leaf-neighbor of one of the $N_{2,1}$ vertices $j$ with $(k_j,\ell_j)=(2,1)$, (B) depicts direct attachment to one of those $N_{2,1}$ vertices, (C) depicts nonleaf attachment other than to one of the $N_{2,1}$, and (D) depicts attachment to a leaf that neighbors a nonleaf $j$ with $(k_j,\ell_j)\ne (2,1)$. Colors designate $(k_i,\ell_i)$ as depicted in the lower left box: black for $(1,0)$, orange for $(2,1)$, blue for $(2,0)$, green for $(3,2)$, cyan for $(3,1)$, and red for $(4,3)$. The probability weight in the PA tree for each of these possibilities is depicted in the lower right box; the actual probabilities for the PA tree are obtained by dividing by the total degree $\sum_{j}k_j=\sum_{k\ge 1}kN_{k}=2(N-1)$.}
    \label{fig:N21_schematic}
\end{figure*}

Vertices of degree $k=2$ have leafdegree $\ell=0$ or $\ell=1$. The only exception occurs for the graph with $N=3$ vertices when the central vertex has $k=\ell=2$. We thus consider the evolution in the $N\geq 4$ range when $N_{2,2}=0$. We limit considerations to $N_{2,1}$, noting $N_{2,0}=N_2-N_{2,1}$. All vertices that become nonleaves pass through the state of having degree $2$ and leafdegree $1$. The stochastic evolution equation
\begin{equation}
\label{N2N1}
N_{2,1}(N+1)=
\begin{cases}
N_{2,1}-1  & \text{prob}~~\frac{N_{2,1}}{N-1}, \\
N_{2,1}+1 & \text{prob}~~\frac{N_1-N_{2,1}}{2(N-1)},\\
N_{2,1}     & \text{prob}~~1-\frac{N_1+N_{2,1}}{2(N-1)}
\end{cases}
\end{equation}
is derived similarly to \eqref{N2N}; see Fig.~\ref{fig:N21_schematic}. Averaging \eqref{N2N1}, we find
\begin{equation}
\label{N21:av-rec}
\langle N_{2,1}(N+1) \rangle=\left(1-\frac{3}{2(N-1)}\right)\langle N_{2,1}\rangle +\frac{\langle N_1\rangle}{2(N-1)}.
\end{equation}

Solving the recurrence \eqref{N21:av-rec} subject to $\langle N_{2,1}(4)\rangle=1$ gives, for $N\geq 4$,
\begin{equation}
\begin{aligned}
\label{eq:N21}
\langle N_{2,1}\rangle
=\frac{2(N-1)}{15}+ \frac{2(N-\frac{2}{5}) \Gamma(N-\frac{5}{2})}{3\sqrt{\pi}\Gamma(N-1)}.
\end{aligned}
\end{equation}
Before obtaining $\langle N_{2,1}^2\rangle$, we obtain the prerequisite correlator $\langle N_{1}N_{2,1}\rangle$. We first write the stochastic form by examining how $(N_1,N_{2,1})$ jointly evolve (see Fig.~\ref{fig:N21_schematic}). Namely,
\begin{equation}
\begin{aligned}
&N_1(N+1)N_{2,1}(N+1)
=\\
&\begin{cases}
(N_1+1)(N_{2,1}-1) & \text{prob}~~ \frac{N_{2,1}}{N-1}, \\
N_1(N_{2,1}+1)      & \text{prob}~~ \frac{N_1-N_{2,1}}{2(N-1)},\\
(N_1+1)N_{2,1}       & \text{prob}~~ 1- \frac{N_1+2N_{2,1}}{2(N-1)},\\
N_1N_{2,1} & \mathrm{otherwise}.
\end{cases}
\end{aligned}
\end{equation}

Averaging yields a recurrence for the correlator \begin{eqnarray}
\langle N_{1}(N+1)N_{2,1}(N+1)\rangle
&=&\left(1-\frac{2}{N-1}\right)\langle N_1N_{2,1}\rangle \nonumber \\
+\left(1-\frac{1}{N-1}\right)\langle N_{2,1}\rangle&+&\frac{\langle N_1^2\rangle}{2(N-1)}. 
\end{eqnarray}
The boundary condition is $\langle N_1(4) N_{2,1}(4)\rangle=2$, from the $\frac{1}{2}$ probability of the line graph $L_4$ wherein $N_1=2$ and $N_{2,1}=2$ (otherwise we have a star graph and $N_{2,1}=0$). We solve to obtain
\begin{equation}
\begin{aligned}
    \langle N_{1}N_{2,1}\rangle &=\frac{(8N-13)(N-1)}{90}\\
    &+\frac{(28 N^2-60N+38) \Gamma(N-\frac{5}{2})}{45\sqrt{\pi}
  \Gamma(N-1)},
\end{aligned}
\end{equation}
from which the covariance is
\begin{eqnarray}
\label{eq:N1N21}
\langle\!\langle N_1N_{2,1}\rangle\!\rangle &=&     \langle N_{1}N_{2,1}\rangle- \langle N_{1}\rangle \langle N_{2,1}\rangle\\
=-\frac{N-1}{18}&-&\frac{16 ( 5 N-2)\Gamma(N-\frac{3}{2})^2 }{\pi \Gamma(N-1)^2}-\frac{8\Gamma(N-\frac{3}{2})}{\sqrt{\pi} \Gamma(N-1)}.\nonumber
\end{eqnarray}
The intensive covariance parameter is thus
\begin{equation}
    \sigma_{1,(2,1)}=-\frac{1}{18}.
\end{equation} 

\ \\

Similarly, we derive a recurrence for $\langle N_{2,1}^2\rangle$ which depends on the equation above for $\langle N_1N_{2,1}\rangle$ and also on the previously derived $\langle N_{1}\rangle$ and $\langle N_{2,1}\rangle$. We have that the square of $N_{2,1}(N+1)$ behaves, by Eq.~\eqref{N2N1}, on average, as
\begin{equation}
\begin{aligned}
\label{N2121:rec2}
\langle N_{2,1}^2(N+1)\rangle &= \left(1-\frac{3}{N-1}\right)\langle N_{2,1}^2\rangle\\
&+\frac{\langle N_1N_{2,1}\rangle}{N-1} +\frac{\langle N_{2,1}\rangle+\langle N_1\rangle}{2(N-1)}.
\end{aligned}
\end{equation}
Using Eqs.~\eqref{eq:N1}, \eqref{eq:N21}, and \eqref{eq:N1N21} for $\langle N_1\rangle$, $\langle N_{2,1}\rangle$, and $\langle N_1 N_{2,1}\rangle$, we can solve the above provided a boundary condition: $\langle N_{2,1}^2(4)\rangle=2$, from the $\frac{1}{2}$ probability of $G_4=L_4$ (the line graph $L_N$ of size $N=4$).
 The solution is
\begin{equation}\begin{aligned}
\langle N_{2,1}^2\rangle&=\frac{ 4 (N-1)^2}{225}+\frac{ 147(N-1)}{1800}\\
&+\frac{(
   40N^2+ 34 N-20)\Gamma(N-\frac{5}{2})}{225 \sqrt{\pi}
   \Gamma(N-1)},
\end{aligned}\end{equation}
from which the variance is
\begin{eqnarray}
&&\langle N_{2,1}^2\rangle-\langle N_{2,1}\rangle^2=\frac{49(N-1)}{600}\\&-&
\frac{ 4 (5 N-2)^2\Gamma(N-\frac{5}{2})^2}{225 \pi \Gamma(N-1)^2}
    +\frac{  18 (5 N-2) \Gamma(N-\frac{5}{2})}{225 \sqrt{\pi} \Gamma(N-1)},\nonumber
\end{eqnarray}
and the associated intensive parameter is 
\begin{equation}\begin{aligned}
\sigma_{(2,1),(2,1)}=\frac{49}{600}.
\end{aligned}\end{equation}

We thus anticipate
\begin{equation}\begin{aligned}
\label{Gauss-21}
P_N(N_{2,1}) \simeq 
 \frac{10}{7}\sqrt{\frac{3}{N\pi}}\exp\!\left\{-\frac{300(N_{2,1}-\frac{2}{15}N)^2}{49 N}\right\},
\end{aligned}\end{equation}
as well as joint normality of $(N_1,N_{2,1})$ with intensive covariance matrix
\begin{equation}
\label{Sigma-21}
\Sigma_{1,(2,1)} = \left[\begin{array}{cc} \sigma_{1,1}& \sigma_{1,(2,1)} \\ \sigma_{1,(2,1)}  & \sigma_{(2,1),(2,1)} \end{array}\right]=\renewcommand{\arraystretch}{1.5}
\left[\begin{array}{cc} \frac{1}{9} & -\frac{1}{18}\\ -\frac{1}{18} & \frac{49}{600}\end{array}\right].
\end{equation}
The joint distribution we conjecture has to leading order a bivariate Gaussian form
\begin{equation}
\label{Gauss-chi}
P_N(N_1,N_{2,1}) \simeq \frac{\exp\!\left[-\frac{1}{2N}\langle \nu| \Sigma_{1,(2,1)}^{-1}|\nu\rangle\right]}{2\pi N \sqrt{\text{det}(\Sigma_{1,(2,1)})}},
\end{equation}
with $\langle \nu|=[N_1-\frac{2}{3}N,N_{2,1}-\frac{2}{15}N]$. Computing the determinant and the inverse of the matrix $\Sigma_{1,(2,1)}$ given by \eqref{Sigma-21} we arrive at 
\begin{equation}\begin{aligned}
P_N(N_1,N_{2,1})
 \simeq \frac{45\sqrt{2}}{\pi N \sqrt{97}}\,\exp\!\left[-\frac{\Psi(N_1,N_{2,1})}{2N}\right],
\end{aligned}\end{equation}
with
\begin{equation}\begin{aligned}
\Psi(N_1,N_{2,1})
&=\frac{1323}{97}\left(N_1-\frac{2}{3}N\right)^2\\
&+\frac{1800}{97}\left(N_1-\frac{2}{3}N\right)\left(N_{2,1}-\frac{2}{15}N\right)\\
&+\frac{1800}{97}\left(N_{2,1}-\frac{2}{15}N\right)^2.
\end{aligned}\end{equation}

We anticipate analogous multivariate normality of arbitrary collections of degree-leafdegree counts $N_{k,\ell}(G)$. Similar results have characterized degree statistics in PA trees \cite{bollobas2003mathematical} and the RRT \cite{Janson05}.

\section{Discussion}
\label{discussion}

In this work, we have introduced the joint degree-leafdegree distribution $n_{k,\ell}$ as a useful descriptor of sparse graphs, and have provided methods for obtaining $n_{k,\ell}$ via multivariate generating function analyses of recursions in the setting of growing random trees. In particular, we have analyzed $n_{k,\ell}$ for the classic PA tree~\cite{barabasi99} and RRT \cite{KRB}. These quantities may be examined in wider model families and real-world data. Indeed, any tree growth model, sparse random graph model, or sparse real-world graph dataset may admit nontrivial joint degree-leafdegree statistics. Quantities similar to $n_{k,\ell}$ have appeared in some past works. One example is the the isotropic complete redirection model \cite{KR17}, where for $k>1$, quantities $c_{k,\ell}$ were defined as the limiting fraction of {\it nonleaves} with degree $k$ and leafdegree $\ell$. There, an exponent $\mu\approx 0.566$ characterizes the anomalous sublinear growth of nonleaves: $N-N_1\sim N^{\mu}$; it was shown that $\mu=\langle \ell/k\rangle$ when $\ell,k$ are drawn with probability $c_{k,\ell}$. A second example is the recent preprint Ref.~\cite{QPA}, in which a nonlinear redirection-based \cite{KR24} model was introduced; in a limiting parameter regime, a recurrence was obtained for $n_{k,\ell}$, albeit differently indexed. We have examined that model in Appendix~\ref{app:crmodel}, analyzing the joint distribution after first demonstrating that a purely leafdegree-based description is tractably applicable.

Our main focus has been on the PA tree, which has been studied extensively \cite{bollobas2003mathematical}. The earliest works characterized the degree distribution, obtaining the cubic powerlaw tail $n_k\sim 4k^{-3}$ \cite{barabasi99,https://doi.org/10.1002/rsa.1009}. Many properties beyond degree have since been characterized, such as the diameter \cite{bollobas2004diameter}, eigenvalue spectrum \cite{arizmendi2022barabasi}, properties of vertices by distance to root \cite{Katona_2005}, nearest neighbor degree correlations \cite{KR01}, and beyond. Numerous extensions to PA have also been considered, typically differing by adjustment of the preference function; for instance, via an additive shift \cite{Sergey00}, nonlinearity \cite{KR01}, age-dependence \cite{WU2014650}, individualized fitnesses \cite{Bianconi_2001}, and triadic closure \cite{holme2002growing}. This work augments those previous characterizations of PA by obtaining the joint distribution $n_{k,\ell}$, \eqref{eq:nkl_PA}, the leafdegree distribution $m_{\ell}$, \eqref{eq:ml_PA}, the protected fraction $p$, \eqref{eq:p_pa}, the degree-stratified protected fractions $n_{k,0}$, \eqref{eq:nk0}, fluctuation results on $N_1$, $N_2$, and $N_{2,1}$, and multivariate combinations thereof (Sec.~\ref{fluctuations}). We have chosen to focus on PA not only to advance the aforementioned literature but also as a case study for techniques of wider applicability. These results from $g(y,z)$ and the joint distribution $n_{k,\ell}$ have provided exact values of numerous properties of interest in PA that are more coarse-grained than $n_{k,\ell}$, namely, $\{m_\ell\}_{\ell\ge 0}$, $p$, and $\{n_{k,0}\}_{k\ge 2}$; we are not aware of more direct methods to obtain these in PA.

Numerous extensions of our techniques are possible in relation to quantities other than leaves, and in data structures beyond trees. For instance, in growth models where new arrivals attach to $m>1$ preexisting vertices, degree-$m$ vertices are analogous to leaves; one can track the statistics of the number of degree-$m$ neighbors a given node has. Extensions to higher-order networks \cite{bianconi2021higher}, e.g., to hypergraphs and simplicial complexes, are also feasible; one can use a notion of `leaves' for hypergraphs as the vertices involved in only one hyperedge. Note that even under the most restrictive definition of leaves (degree-$1$ vertices with a single dyadic edge), higher-order networks may exhibit nontrivial leaf statistics. Of particular interest would be the leaf statistics of higher-order growth models \cite{PK-hyper,oh2024exploring}. A second direction that we emphasize is based on how descriptive analyses of the type pursued herein can allow corresponding formulation of models with new mechanisms---namely, those mechanisms tractable under the introduced descriptive tools. For example, leafdegree-based shifted linear PA \cite{hartle2025statistics}, which exhibits a wider range of behaviors than classic degree-based shifted linear PA, was introduced after the leafdegree distribution became an object of interest~\cite{hartle2025growing}.

Closer to this particular work would be the analysis of $n_{k,\ell}$ in related models. Shifted linear preferential attachment models \cite{Simon55} with attachment rate proportional to $k+\delta$ constitute a one-parameter family of such models. The leafdegree distribution does not satisfy a closed recurrence, as with the PA tree considered herein ($\delta=0$). Similarly, the recently introduced leaf-based preferential attachment model (attach to $j$ with probability $\propto \ell_j+a$ for some $a>0$) \cite{hartle2025statistics} doesn't admit a closed recurrence for its degree distribution $n_k$. Yet we anticipate that the joint distribution $n_{k,\ell}$ would be tractable, from which $n_k$ could be obtained. Special cases of interest include $a=1$, where $m_\ell\sim e^{-2\sqrt{ g \ell}}$ with $g$ the golden ratio, and $a = 1 - 1/\sqrt{3}$, where $m_\ell\sim \ell^{-3}$. The same situation arises in the model considered in Appendix~\ref{app:crmodel}; there is a closed recurrence for $n_{k,\ell}$, but a lack thereof for $n_k$. Also of interest is the joint degree-leafdegree distribution in unlabeled random graphs, including growth models \cite{hartle2025growing} and static ensembles \cite{SCHWENK197771,luczak1991deal,paton2022entropy,evnin2025ensemble}.

More broadly, we anticipate use cases of the methods developed herein, and leaf-based descriptions and model mechanisms, in random graph theory and in real-world graph data modeling. Leaves often play a distinguished functional role in real world networks, such as sources and consumers in flow networks \cite{west} and observed species in phylogenetic trees \cite{rieux}. The statistics of leaves are readily measurable in data; of interest are (i) modeling via conventional models (e.g., degree-based preferential attachment) using leaf-related test statistics---comparing to model theoretical values obtained with approaches like those herein---e.g., testing $\bar{\ell}(k)$ in real vs model networks, but also (ii) modeling of real networks with leaf-based statistical mechanisms, such as leafdegree-based preferential attachment~\cite{hartle2025statistics}, the redirection model of Ref.~\cite{QPA} considered in Appendix~\ref{app:crmodel}, unlabeled growth of leaf-symmetric trees~\cite{hartle2025growing}, and many other possible mechanisms yet to be examined.

\appendix

\section{Random recursive tree}
\label{rrt}

In this Appendix, we analyze the joint distribution of degree and leafdegree in random recursive trees (RRTs). RRTs are generated via the simplest stochastic growth procedure: vertices are added one by one, and each new vertex attaches to a uniformly randomly chosen existing vertex \cite{Newman,book,Frieze,Hofstad,DM22}. As with our treatment of PA, we obtain recursions for $n_{k,\ell}$ and the associated PDE for $g(y,z)$, which we solve, subsequently obtaining numerous results through analysis of $g(y,z)$.

For the RRT, each existing vertex is attached to with probability $1/N$, so a degree-$k$ leafdegree-$\ell$ vertex is attached to with probability $N_{k,\ell}/N$, and a leaf-neighbor thereof is attached to with probability $\ell N_{k,\ell}/N$. (Indeed, this recovers the total leaf-attachment probability: $\sum_{0\le \ell<k<\infty}\ell N_{k,\ell}/N=N_1/N$.) Furthermore, one new leaf arrives at each attachment. Thus $N_{k,\ell}$ varies as
\begin{equation}\begin{aligned}
\frac{dN_{k,\ell}}{dN}&=\frac{N_{k-1,\ell-1}}{N}-\frac{N_{k,\ell}}{N}+\frac{(\ell+1)N_{k,\ell+1}}{N}\\
&-\frac{\ell N_{k,\ell}}{N}+\delta_{k,1}\delta_{\ell,0},
\end{aligned}\end{equation}
from which we deduce the recurrence for $n_{k,\ell}$:
\begin{equation}
\label{n:rec-RRT}
(\ell+2)n_{k,\ell} = n_{k-1,\ell-1}+(\ell+1)n_{k,\ell+1}+\delta_{k,1}\delta_{\ell,0}. 
\end{equation}
Summing over $\ell$ we recover the recurrence  for the degree distribution, $2n_k=n_{k-1} +\delta_{k,1}$, which admits the well-known solution $n_k = 2^{-k}$. Remarkably, the summation over $k$ also leads to a closed recurrence \cite{hartle2025statistics}
\begin{equation}
(\ell+2) m_\ell = m_{\ell-1}+(\ell+1) m_{\ell+1}+\delta_{\ell,0},
\end{equation}
which admits the solution
\begin{equation}
m_\ell = \frac{\gamma(\ell +1, 1)}{\Gamma(\ell +1)}.
\end{equation}

To determine the joint degree distribution, we use again the generating function (Eq.~\eqref{eq:gdef}) and recast the recurrence \eqref{n:rec-RRT} into a PDE
\begin{equation}
\label{n-yz:RRT}
(1-z)\,\frac{\partial g}{\partial z}-(2-yz)g+y=0.
\end{equation}
We seek solution in the form
\begin{equation}
\label{f:def}
g(y,z) = (1-z)^{y-2} e^{yz} f(y,z),
\end{equation}
The integrating factor solves the homogeneous version of \eqref{n-yz:RRT}, and therefore the ansatz \eqref{f:def} reduces \eqref{n-yz:RRT} to
\begin{equation}
\label{f:eq}
\frac{\partial f}{\partial z} = -y(1-z)^{1-y} e^{-yz},
\end{equation}
Integrating \eqref{f:eq} gives
\begin{equation*}
g(y,z) =  y(1-z)^{y-2} e^{yz} \int_z^1 dw\,(1-w)^{1-y} e^{-yw},
\end{equation*}
which we transform to
\begin{equation}
\label{n-yz:sol}
g(y,z) =  y \int_0^1 (1-t)^{1-y} e^{-y(1-z)t}dt.
\end{equation}

Equivalently, with change of integration variable to $v=1-t$,
\begin{equation}\begin{aligned}
g(y,z) =  y \int_0^1 v^{1-y} e^{-y(1-z)(1-v)}dv,
\end{aligned}\end{equation}
as reported in Eq.~\eqref{rrt_g}. One may verify the normalization and moment conditions
\begin{equation}\begin{aligned}
g(1,1)&=1,\\
\partial_yg(1,1)&=2,\\
\partial_yg(1,1)&=1/2.
\end{aligned}\end{equation}

\subsection{$n_{k}(z)$ and $n_{k,\ell}$}

We now proceed to derive an integral expression for $n_{k,\ell}$ by expansion of $n_{k}(z)$ in powers of $z$, after obtaining $n_{k}(z)$ by expansion of $g(y,z)$ in powers of $y$. We write
\begin{equation}\begin{aligned}
g(y,z)=y\int_0^1e^{-y[(1-v)(1-z)+\log v]}v\,dv
\end{aligned}\end{equation}
and expand it to find
\begin{equation}
\label{nkz:RRT}
n_k(z) = \int_0^1\frac{[-(1-v)(1-z)-\log v]^{k-1}}{(k-1)!}\,v\,dv.
\end{equation}
Expanding in powers of $z$ we arrive at
\begin{equation}
\label{eq:nkl-rrt}
n_{k,\ell}=\int_0^1\frac{(1-v)^\ell [v-1-\log v]^{k-1-\ell}}{\ell!(k-1-\ell)!}\,v\,dv.
\end{equation}
The fractions $n_{k,\ell}$ are rational numbers. This property is not immediately obvious from the exact formula Eq.~\eqref{eq:nkl-rrt}, yet it follows from the structure of the recurrence \eqref{n:rec-RRT}.

\subsection{Special cases}

Specializing Eq.~\eqref{eq:nkl-rrt} to $\ell=k-1$ we deduce the fraction of vertices that are near-stars 
\begin{equation}\begin{aligned}
n_{k,k-1}=\frac{1}{(k+1)!}\,.
\end{aligned}\end{equation}
The total fraction of vertices that are near-stars is
\begin{equation}\begin{aligned}
\sum_{k=1}^{\infty}n_{k,k-1}=e-2=0.718\, 281\, \ldots
\end{aligned}\end{equation}
Similarly, we compute
\begin{equation}
n_{k,k-2} = \frac{H_k}{k!}-\frac{2k}{(k+1)!}
\end{equation}
where $H_k$ is the $k$th harmonic number.

The mean leafdegree $\bar\ell(k)$ of vertices with fixed degree $k$ is given by the genearal formula \eqref{mean-leaf}, see also \eqref{mean-leaf-2}, which becomes $\bar{\ell}(k)=2^kn'_k(z)|_{z=1}$ for the RRT. Using \eqref{nkz:RRT} we arrive at 
\begin{equation}\begin{aligned}
\bar{\ell}(k)&=\frac{2^k}{(k-2)!}\int_0^1\left(-\log v\right)^{k-2}\left(1-v\right)vdv\\
&=2-3\left(\frac{2}{3}\right)^k,
\end{aligned}\end{equation}
which saturates in the $k\to\infty$ limit; a vertex of large degree has on average two leaves among its neighbors. This is in contrast with PA, where $\bar{\ell}(k)\simeq \frac{k}{2}$ in the large $k$ limit. The conditional distribution of leafdegree, viz., the quantity
\begin{equation}
\label{stationary}
P_\infty(\ell) = \lim_{k\to\infty}\frac{n_{k,\ell}}{n_k}\,,
\end{equation}
is the Poisson distribution with mean $2$:
\begin{equation}
\label{Poisson}
P_\infty(\ell) = \frac{2^\ell e^{-2}}{\ell!}\,. 
\end{equation}
One can derive \eqref{Poisson} starting with the exact integral representation \eqref{eq:nkl-rrt} for $n_{k,\ell}$ and extracting the behavior in the $k\to\infty$ limit with $\ell$ kept fixed. Another derivation is based on a direct evaluation of the $q$th moment $[d^q n_k(z)/dz^q]_{z=1}$ and showing that it converges to $2^q$ in the $k\to\infty$ limit. An alternative derivation is based on substituting \eqref{stationary} into \eqref{n:rec-RRT}, recalling $n_k=2^{-k}$, and arriving at a recurrence
\begin{equation}
\label{ell:rec-RRT}
(\ell+2)P_\infty(\ell) = 2P_\infty(\ell-1)+(\ell+1)P_\infty(\ell+1)
\end{equation}
which is solved to yield $e^{-2}2^{\ell}/\ell!$, with the amplitude fixed by normalization. This differs from $m_\ell=\gamma(\ell+1,1)/\ell!$, with the latter accounting for all vertices, most of which are not high degree.

\subsection{Protected vertices in RRTs}

An immediacy from $g(y,z)$ is the asymptotic fraction of vertices with leafdegree zero:
\begin{equation}\begin{aligned}
m_0=g(1,0)=  e^{-1}\int_0^1  e^{v}dv=1-\frac{1}{e},
\end{aligned}\end{equation}
recovering the well-known protected fraction $p=\frac{1}{2}-\frac{1}{e}$. Also of interest is the degree-stratified fractions of protected vertices; for $k\ge 2$, we have
\begin{equation}\begin{aligned}
\label{eq:nk0-rrt}
n_{k,0}= \frac{1}{(k-1)!}\int_0^1  (v-1-\log v)^{k-1}vdv,
\end{aligned}\end{equation}
which also reproduces $n_{1,0}=n_1=1/2$. To determine the large $k$ asymptotic we change the variable, $v=e^{-u}$ so that $dv=-e^{-u}du$, $v=0\Rightarrow u=\infty$, $v=1\Rightarrow u=0$, and rewrite Eq.~\eqref{eq:nk0-rrt} as 
\begin{equation}\begin{aligned}
n_{k,0}&=\frac{1}{(k-1)!}\int_0^\infty  (u-1+e^{-u})^{k-1}e^{-2u}du\\
&\simeq \frac{1}{(k-1)!}  \int_1^\infty du\,e^{-2u}(u-1)^{k-1}= \frac{e^{-2}}{2^k},
\end{aligned}\end{equation}
also leading to
\begin{equation}\begin{aligned}
\lim_{k\uparrow \infty} \frac{n_{k,0}}{n_k} = e^{-2} = 0.135\, 335\, \ldots;
\end{aligned}\end{equation}
thus only a slightly larger fraction of vertices at $k\gg 1$ are protected than the overall protected fraction; the deficit is $e^{-2}-p\approx 0.003\, 214$. We can furthermore obtain the average degree of protected vertices in RRTs:
\begin{equation}\begin{aligned}
\bar{k}_{p}&=\frac{\partial_y g(1,0)-\frac{1}{2}}{\frac{1}{2}-\frac{1}{e}} 
=\frac{\mathrm{Ei}(1)-\gamma}{\frac{e}{2}-1}-1\\
&=2.669\, 596\, 248\, \ldots\\\
\end{aligned}\end{equation}
This is notably larger than the average degree of protected vertices in PA, $\bar{k}_p\approx 2.20$ (Eq.~\eqref{eq:kp_PA}).

\subsection{Correlations}
\label{sub:corr-RRT}

To compute $\langle k\ell\rangle$, we use the formula
\begin{equation}
\langle k\ell\rangle=\left.\frac{\partial^2 g(y,z)}{\partial y\partial z}\right\vert_{y=z=1}.
\end{equation}
From the integral representation of $g$ we have
\begin{equation*}
\frac{\partial^2 g(y,z)}{\partial y\partial z}=\frac{\partial^2}{\partial y\partial z}\left[y\int_0^1 v^{1-y}e^{-y(1-z)(1-v)} dv\right]
\end{equation*}
Evaluating the derivative at $y=z=1$, we have
\begin{equation}
\langle k\ell\rangle=\int_0^1 (2-\log v)(1-v)dv=\frac{7}{4}\,.
\end{equation}
Compare to $\langle k\rangle\langle\ell\rangle=2\times \frac{1}{2} = 1$.

\section{Derivation of \eqref{Gauss-1}}
\label{ap:Gauss}

We write $F(N_1,N)\equiv P(N_1; N+1)$, so \eqref{PN1N} becomes
\begin{eqnarray}
\label{FN}
F(N_1, N+1) &=& \left(1-\frac{N_1-1}{2N}\right)F(N_1-1, N)  \nonumber \\
&+&\frac{N_1}{2N}\,F(N_1, N).
\end{eqnarray}
We treat $N\gg 1$ and $N_1\gg 1$ as continuous variables and expand \eqref{FN} in Taylor series up to the second derivatives (which is sufficient as it will soon become clear) 
\begin{eqnarray*}
\frac{\partial F}{\partial N}+\frac{1}{2}  \frac{\partial^2 F}{\partial N^2} = 
\left(1-\frac{N_1}{2N}\right) \left(-\frac{\partial F}{\partial N_1}+\frac{1}{2}  \frac{\partial^2 F}{\partial N_1^2}\right) + \frac{F}{2N}  .
\end{eqnarray*}
Instead of $N_1$, we use the variable
\begin{equation}
\label{xi}
\xi = \frac{N_1-\frac{2}{3}N}{\sqrt{N}}.
\end{equation}
In these new variables, the function $\Phi(\xi, N)=F(N_1,N)$ satisfies in the leading order an equation 
\begin{equation}
\label{Phi}
N\frac{\partial \Phi}{\partial N}=\xi\frac{\partial \Phi}{\partial \xi}+\frac{1}{9}  \frac{\partial^2 \Phi}{\partial \xi^2}+\frac{1}{2}\Phi.
\end{equation}
In deriving \eqref{Phi}, we have used  
\begin{equation*}
\frac{\partial }{\partial N}\to \frac{\partial }{\partial N}-\left(\frac{2}{3\sqrt{N}}+\frac{\xi}{2N}\right)\frac{\partial }{\partial \xi}\,,
\quad \frac{\partial }{\partial N_1}\to \frac{1}{\sqrt{N}}\,\frac{\partial }{\partial \xi}.
\end{equation*}
The solution to \eqref{Phi} is $\Phi = C N^{-1/2}e^{-9\xi^2/2}$. Recalling \eqref{xi} and fixing the amplitude $C$ to ensure normalization we arrive at \eqref{Gauss-1}. 

\section{Cumulants of $N_1$ and $N_2$}
\label{ap:cum}

To derive the cumulant generating function \eqref{cum:leaves} of the number of leaves $N_1$, we use the stochastic evolution equation \eqref{N1N} for this quantity and find
\begin{equation}
\label{N1-cum}
\left\langle e^{x N_1(N+1)}\right\rangle = e^x\left\langle e^{x N_1}\right\rangle+\frac{1-e^x}{2(N-1)}\left\langle N_1 e^{x N_1}\right\rangle.
\end{equation}
Hereinafter we write again $N_1$ instead of $N_1(N)$; we use the complete notation only when the number of vertices is $N+1$. It proves convenient to rewrite \eqref{N1-cum} as
\begin{equation}
\label{log:eq}
\left\langle e^{x N_1(N+1)}\right\rangle = e^x\left\langle e^{x N_1}\right\rangle+\frac{1-e^x}{2(N-1)}\frac{d \left\langle e^{x N_1}\right\rangle}{d x}.
\end{equation}
The first two cumulants, the average and the variance, have the leading terms proportional to $N-1$, and subleading terms vanishing as $N^{-1/2}$ in the large $N$ limit, cf. \eqref{eq:N1} and \eqref{N1:var}. Therefore we define the (normalized) cumulant generating function via 
\begin{equation}
\label{Q:def}
Q(x) = \lim_{N\to\infty} \frac{\log \left\langle e^{x N_1}\right\rangle}{N-1},
\end{equation}
we recast \eqref{log:eq} into an ordinary differential equation (ODE)
\begin{equation}
\label{Q:eq}
e^{Q} = e^x +\frac{1-e^x}{2}\,\frac{d Q}{d x}.
\end{equation}
The transformation $u(x)= e^{-Q(x)}$ recasts \eqref{Q:eq} into a linear (inhomogeneous) differential equation which we solve subject to $u(0)=1$ corresponding to $Q(0)=0$. Returning back to $Q(x)$, we arrive at 
\begin{equation}
\label{Q:sol}
Q(x) = \log\,\frac{(e^x-1)^2}{2(e^x-1-x)},
\end{equation}
leading to the announced result \eqref{cum:leaves}.

We now outline the computation of the cumulants of random variables $N_1$ and $N_2$. The stochastic evolution equation for the pair $(N_1,N_2)$ leads to
\begin{eqnarray}
\label{log:xy}
\left\langle e^{x N_1(N+1)+yN_2(N+1)}\right\rangle &=& \left\langle \frac{N_2}{N-1}\,e^{x(N_1+1)+y(N_2-1)}\right\rangle \nonumber \\
&+& \left\langle \frac{N_1}{2(N-1)}\,e^{x N_1+y(N_2+1)}\right\rangle  \nonumber  \\
&-& \left\langle \frac{N_1+2N_2}{2(N-1)}\,e^{x (N_1+1)+y N_2}\right\rangle  \nonumber  \\
&+& \left\langle e^{x (N_1+1)+y N_2}\right\rangle. 
\end{eqnarray}
We rewrite the right-hand side (RHS) of Eq.~\eqref{log:xy} as
\begin{eqnarray}
\label{RHS}
\text{RHS} &=& e^x \left\langle e^{x N_1+y N_2}\right\rangle 
+ \frac{e^{x-y}-e^x}{N-1}\,\frac{\partial}{\partial y} \left\langle e^{x N_1+y N_2}\right\rangle \nonumber  \\
&+&\frac{e^y-e^x}{2}\,\frac{\partial }{\partial x}  \left\langle e^{x N_1+y N_2}\right\rangle,
\end{eqnarray}
and use the two-variate cumulant generating function 
\begin{equation}
\label{Qxy}
Q(x,y) = \lim_{N\to\infty} \frac{\log \left\langle e^{x N_1+yN_2}\right\rangle}{N-1}.
\end{equation}
to reduce \eqref{log:xy}--\eqref{RHS} to a non-linear PDE
\begin{equation}
\label{Q:PDE}
e^{Q} = e^x +\frac{e^y-e^x}{2}\,\frac{\partial Q}{\partial x} +\big(e^{x-y}-e^x\big)\,\frac{\partial Q}{\partial y}.
\end{equation}
Making the transformation
\begin{equation}
u(x,y) = e^{-Q(x,y)},
\end{equation}
we recast \eqref{Q:PDE} into a linear inhomogeneous PDE
\begin{equation}
\label{u:PDE}
\frac{e^{y-x}-1}{2}\,\frac{\partial u}{\partial x} +\big(e^{-y}-1\big)\,\frac{\partial u}{\partial y} + e^{-x} = u .
\end{equation}

First order PDEs are hyperbolic and solvable by the method of characteristics. In the present, it is convenient to begin with a transformation
\begin{equation}
\label{eta}
\eta=\log(e^y-1),  \qquad v(x,\eta) = e^\eta u(x,y),
\end{equation}
recasting \eqref{u:PDE} into 
\begin{equation}
\label{v:PDE}
\frac{1-e^{-x}(e^\eta+1)}{2}\,\frac{\partial v}{\partial x} +\frac{\partial v}{\partial \eta} = e^{\eta -x} .
\end{equation}
The characteristics are now defined by an ODE
\begin{equation}
\label{x:char-eq}
\frac{dx}{d\eta}=\frac{1-e^{-x}(e^\eta+1)}{2},
\end{equation}
which is integrated to give
\begin{equation}
\label{x:char}
e^x = 1 - e^\eta + 2Ce^\frac{\eta}{2}.
\end{equation}
Characteristics are parametrized by $C$. Along each characteristic, Eq.~\eqref{v:PDE} becomes an ODE
\begin{equation}
\label{v:ODE}
\frac{d v}{d \eta} = e^{\eta -x} = \frac{e^\eta}{1 - e^\eta + 2Ce^\frac{\eta}{2}}.
\end{equation}
Integrating \eqref{v:ODE} yields
\begin{eqnarray*}
v &=& \frac{2C}{\sqrt{C^2+1}}\,\tanh^{-1}\left(\frac{C+e^\frac{\eta}{2}}{\sqrt{C^2+1}}\right) \nonumber \\
&-& \log\!\left[1-e^\eta+2Ce^\frac{\eta}{2}\right] + F(C),
\end{eqnarray*}
with integration `constant' $F(C)$ depending on the characteristic. Requiring that $u = v/(e^y-1)$ does not diverge in the $y\to 0$ limit we fix the integration constant:
\begin{eqnarray}
v &=& 2c\left[\tanh^{-1}\left(\frac{C+e^\frac{\eta}{2}}{\sqrt{C^2+1}}\right)-\tanh^{-1}(c)\right] \nonumber \\
&-& \log\!\left[1-e^\eta+2Ce^\frac{\eta}{2}\right], \ 
\ \ c = \frac{C}{\sqrt{C^2+1}}. \ 
\end{eqnarray}
Returning to the original variables we arrive at the parametric representation of the solution:
\begin{equation}
\label{u:sol}
u = \frac{2c\!\left[\tanh^{-1}\left(\frac{C+\sqrt{e^y-1}}{\sqrt{C^2+1}}\right)-\tanh^{-1} (c)\right] - x}{e^y-1},
\end{equation}
with
\begin{subequations}
\label{Cc}
\begin{align}
C & =\frac{e^x+e^y-2}{2\sqrt{e^y-1}},\\
c & = \frac{2 \sqrt{e^y-1} \left(e^y+e^x-2\right)}{2 e^{x+y}+e^{2 x}+e^{2 y}-4 e^x}.
\end{align}
\end{subequations}

Extracting the cumulants from the exact parametric solution \eqref{u:sol}--\eqref{Cc} is not easy due to an apparent singularity at $(x,y)=(0,0)$. One can employ a pedestrian approach and compute the cumulants of small order by substituting an expansion
\begin{equation}
Q(x,y)=\sum_{p,q\geq 1} \kappa^{(p,q)}_{12}\frac{x^p\,y^q}{p!\,q!}
\end{equation}
into \eqref{Q:PDE}. Equating the first order terms we recover 
\begin{equation*}
\kappa^{(1,0)}_{12} = n_1 = \frac{2}{3}\,, \quad \kappa^{(0,1)}_{12} = n_2 = \frac{1}{6}.
\end{equation*}
Equating  the second order terms we recover the normalized variances and the covariance
\begin{equation*}
\kappa^{(2,0)}_{12}=\frac{1}{9}\,, \quad \kappa^{(1,1)}_{12}=-\frac{4}{45}\,,  \quad \kappa^{(0,2)}_{12}=\frac{23}{180}.
\end{equation*}
In the third order, we recover $\kappa^{(3,0)}_{12}=\kappa_1^{(3)}=-\frac{1}{135}$, and obtain three new normalized cumulants
\begin{equation}
\begin{split}
&\kappa^{(2,1)}_{12}=\lim_{N\to\infty}\frac{\langle\!\langle N_1^2 N_2\rangle\!\rangle}{N} = \frac{7}{270}, \\
&\kappa^{(1,2)}_{12}=\lim_{N\to\infty}\frac{\langle\!\langle N_1 N_2^2\rangle\!\rangle}{N} = -\frac{46}{945}, \\
&\kappa^{(0,3)}_{12}=\kappa_2^{(3)}=\lim_{N\to\infty}\frac{\langle\!\langle N_2^3\rangle\!\rangle}{N} = \frac{13}{189}.
\end{split}
\end{equation}

Continuing up to the $6^\text{th}$ order, we obtain the results collected in Table~\ref{Table:kappa-12}. The first column constitutes the results announced in  Table~\ref{Table:kappa-2}.

\begin{table}[h!]
\centering
\setlength{\tabcolsep}{0.8pt}
\renewcommand{\arraystretch}{1.3}{
\begin{tabular}{| c | c | c | c | c | c | c | c |}
\hline
\diagbox{$q$}{$p$}  & $0$ & $1$ & $2$ & $3$ & $4$ & $5$ & $6$\\ 
\hline
$0$    &  0                    &    $\frac{2}{3}$   &  $\frac{1}{9}$     & $-\frac{1}{135}$        &  $-\frac{2}{135}$ &  $\frac{1}{567}$ & $\frac{68}{8505}$    \\ 
\hline 
$1$   & $\frac{1}{6}$    &  $-\frac{4}{45}$   &  $\frac{7}{270}$   &  $\frac{17}{1890}$    &  $-\frac{337}{28350}$  & $-\frac{71}{17010}$  & \\
\hline 
$2$  & $\frac{23}{180}$ & $-\frac{46}{945}$ &  $-\frac{1}{175}$  & $\frac{43}{2025}$    & $-\frac{151}{85050}$ &  &\\ 
\hline
$3$  & $\frac{13}{189}$    & $\frac{2}{675}$      & $-\frac{1817}{56700}$                       &$\frac{102727}{9355500}$   &  &   &\\ 
\hline
$4$  & $-\frac{251}{37800}$ & $\frac{7466}{155925}$          & $-\frac{388301}{16372125 }$ &   &  &    & \\
\hline
$5$  & $-\frac{445}{6237}$ &     $\frac{1713926}{42567525}$ &   &  &  &   & \\
\hline
$6$  & $-\frac{9790117}{170270100}$ &        &  &   &  &   &   \\
\hline
\end{tabular}
}
\caption{The normalized two-variate cumulants $\kappa_{12}^{(p,q)}$ up to the fifth order, $p+q\leq 6$. The normalized cumulants $\kappa_{12}^{(p,0)}=\kappa_1^{(p)}$ are the cumulants of $N_1$. The normalized cumulants $\kappa_{12}^{(0,q)}=\kappa_2^{(q)}$ are the cumulants of $N_2$. } 
\label{Table:kappa-12}
\end{table}

\section{Constant redirection model}
\label{app:crmodel}

A recent preprint \cite{QPA} considers growing tree models with a redirection mechanism \cite{KR24}. At each growth step, an attachment weight $w_i$ is assigned to each vertex $i$, and the probability of attachment to vertex $i$ is $w_i/\sum_v w_v$, where the sum runs over all vertices $v$. We consider one of the models of Ref.~\cite{QPA}, named {\it constant redirection} (CR), which has parameters $\alpha\in\mathbb{R}$ and $r\in[0,1]$. Mechanistically, the model consists of selecting a random `target' vertex $t$ with probability proportional to $k_t^\alpha$, where $k_t$ is the current degree of vertex $t$. Then with probability $1-r$, the target $t$ is attached to directly; otherwise (probability $r$) the attachment is `redirected' to a random neighbor of $t$. This leads to attachment weight $w_i$ for vertex $i$ given by
\begin{equation}
w_i(\alpha,r)=(1-r)k_i^\alpha+r\sum_{j\sim i}k_{j}^{\alpha-1},
\end{equation}
with $j\sim i$ denoting neighbors $j$ of $i$. The model with extreme nonlinearity parameter $\alpha=-\infty$ at the value $r=\frac{1}{2}$ has interesting properties, and it also arises as a limiting case of another model introduced in Ref.~\cite{QPA}. 

In this Appendix, we show how to analyze the model with $(\alpha, r)=(-\infty, \frac{1}{2})$ using the approaches developed herein and in recent works \cite{hartle2025growing,hartle2025statistics}. We first derive a closed recurrence for the leafdegree $(m_\ell)_{\ell\ge 0}$ and solve it. The recurrence equation for the joint distribution $n_{k,\ell}$ was derived in \cite{QPA}, and we use to deduce several properties of the joint distribution.

\subsection{The limiting case $\alpha=-\infty$, $r=\frac{1}{2}$}

For the CR model with $r=\frac{1}{2}$, we redefine the weight, $w_i\rightarrow 2w_i$. Denoting $w_i(\alpha)=w_i(\alpha,\frac{1}{2})$, we have
\begin{equation}
w_i(\alpha)=k_i^\alpha+\sum_{j\sim i}k_j^{\alpha-1}.
\end{equation}
The attachment probability to node $i$ is then
\begin{equation}
p_i(\alpha)=\frac{w_i(\alpha)}{\sum_{v}w_v(\alpha)}=\frac{k_i^\alpha+\sum_{j\sim i}k_j^{\alpha-1}}{\sum_{v}\left(k_v^\alpha+\sum_{j\sim v}k_j^{\alpha-1}\right)}
\end{equation}
When $\alpha = -\infty$, only terms with $k_i=1$ remain:
\begin{equation}
w_i(-\infty)=\delta_{1,k_i}+\ell_i.
\end{equation}
Summing this $w_i$ over $i$, we get $2N_1$, since each leaf is counted once by each term. Thus
\begin{equation}
p_i = \frac{
\delta_{1,k_i}+\ell_i}{2N_1}.
\end{equation}
Since $k_i=1$ always implies $\ell_i=0$, the attachment probability can also be expressed as
\begin{equation}
\label{eq:pi_cr}
p_i = \frac{1}{2N_1}\times
\begin{cases}
1,      & k_i=1,\\
\ell_i, & k_i>1.
\end{cases}
\end{equation}

Thus $p_i$ vanishes for all $i$ that are {\it protected}: $k_i>1$, $\ell_i=0$. These attachment weights are the same as in the leaf-based preferential attachment model of Ref.~\cite{hartle2025statistics} in the limit $a=0$, with the exception of leaves, which, in this model, are granted weight $1$. As a simple illustration, the attachment weights of vertices with $k\leq 6$ are displayed in Table~\ref{tab:crmodelweights}.

\begin{table}[h!]
\centering
\setlength{\tabcolsep}{3pt}
\renewcommand{\arraystretch}{1.0}{
\begin{tabular}{| c | c | c | c | c | c | c |}
\hline
\diagbox{$k$}{$\ell$}  & $0$ & $1$ & $2$ & $3$ & $4$ & $5$\\ 
\hline
$1$ & 1 &   \    &  \    & \      & \   & \   \\ 
\hline
$2$ & 0 & $1$ &  \    &  \       &  \  & \   \\
\hline
$3$ & 0  & $1$ &  $2$ &\  & \   & \   \\
\hline
$4$ & 0 & $1$ & $2$ & $3$     & \   & \    \\
\hline
$5$ & 0 & $1$ & $2$ & $3$ & $4$  & \   \\
\hline 
$6$ & $0$ & $1$ & $2$ & $3$ & $4$  & $5$   \\
\hline 
\end{tabular}
\caption{The attachment weights for vertices with degree $k_i=k$ and leafdegree $\ell_i=\ell$ in the CR model at $\alpha=-\infty$, $r=\frac{1}{2}$; the attachment probabilities are $\frac{1}{2N_1}\times$ the values displayed. Except distinguishing protected vertices from leaves, the weight depends only on leafdegree $\ell$ (Eq.~\eqref{eq:pi_cr}). Leaves have weight $1$ and nonleaves $i$ have weight $\ell_i$, totaling to $2N_1$. }
\label{tab:crmodelweights}
}
\end{table}

\subsection{Total leaf count}

Due to the dependence of attachment weight only on leafdegree (except at $\ell=0$, see Table~\ref{tab:crmodelweights}), 
we expect a concise nontrivial description of the system in terms of $(n_1,p,m_1,m_2,m_3,\ldots)$. At every arrival beyond $N=3$, the probability of a creating a leaf is exactly $\tfrac{1}{2}$, because a leaf is always targeted. Therefore, the exact distribution of total leaf-count at any $N\ge 3$ is binomial:
\begin{equation}
\mathbb{P}(\, N_1(G_N)=L\, )=\frac{1}{2^{N-3}}\binom{N-3}{L-2},
\end{equation}
implying $n_1=\tfrac{1}{2}$. From this complete description of $N_1(G_N)$ we can directly observe extensivity and self-averaging.

\subsection{Leafdegree distribution}
We next examine leafdegree-stratified counts $(M_{\ell})_{\ell\ge 0}$, where $M_0=N_1+P$. Conjecturing extensivity and self-averaging, we examine the limiting fractions $p,(m_{\ell})_{\ell\ge 0}$. We start in Sec.~\ref{app:crmodel_recu} by obtaining a recurrence that determines $m_\ell$ at arbitrary $\ell\ge 0$ as a function of $p$. The value of $p$ is uniquely determined by normalization; in a later section (Sec.~\ref{app:crmodel_gf}), we obtain $p$ explicitly via generating function \cite{Wilf}, and examine the tail behavior of $m_\ell$ via a WKB approach \cite{Bender}.

\subsubsection{Leafdegree recurrence}
\label{app:crmodel_recu}

\begin{figure*}
\includegraphics[scale=0.33,trim=200 200 0 0]{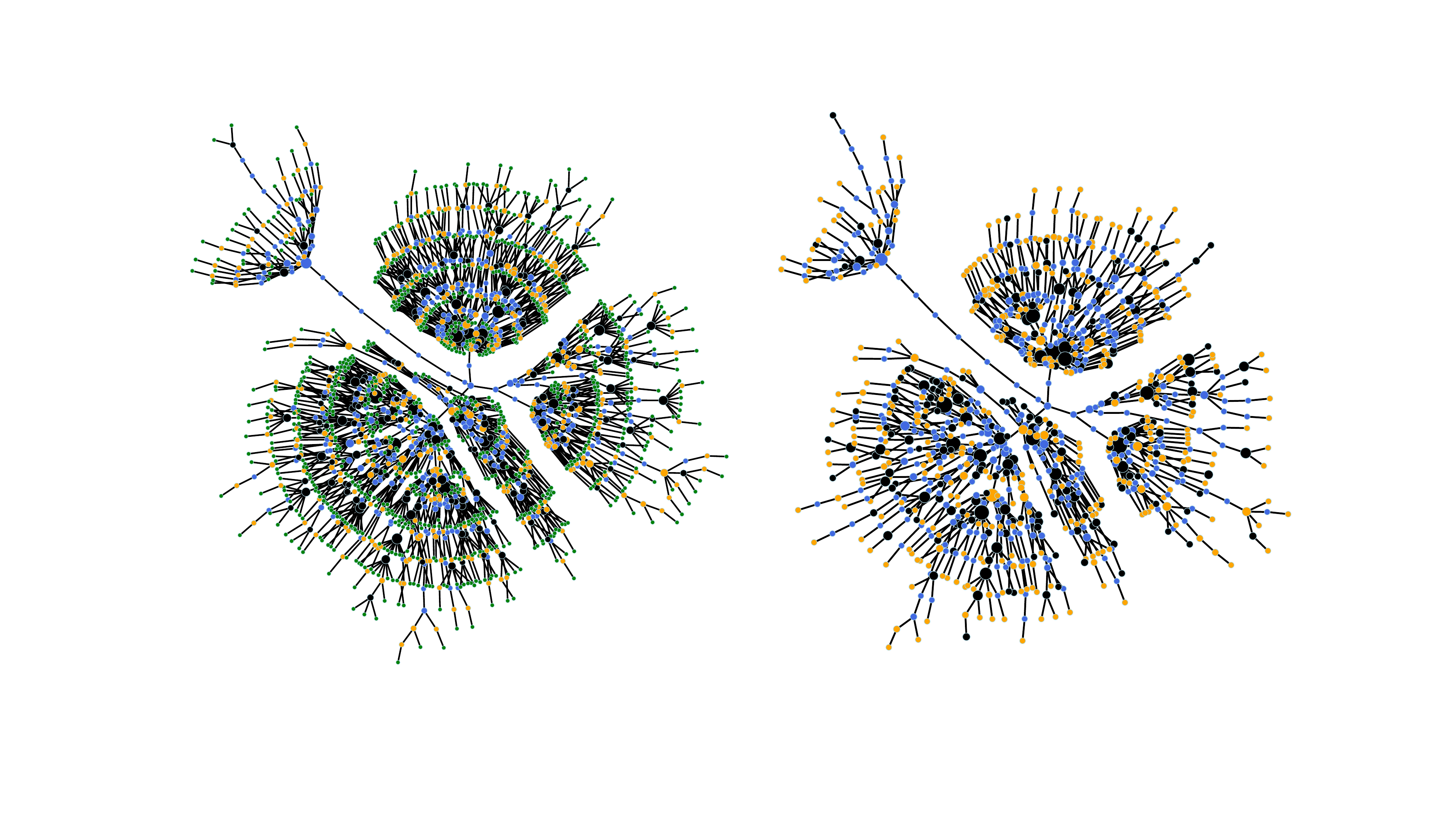}
\caption{A random tree from the CR model with $\alpha=-\infty$, $r=\frac{1}{2}$. Left: the full-sized tree with $N=2000$ vertices; colors distinguish leaves (green, $N_1=1016$), leafdegree-$1$ vertices (orange, $M_1=394$), and protected vertices (blue, $P=386$). Right: the subgraph consisting of all nonleaves, with the same color assignment.}
\label{fig:emb}
\end{figure*}

We now obtain equations for how $P$ and $(M_{\ell})_{\ell\ge 1}$ evolve; we begin with $P$, where $P/N$ converges to the limiting protected fraction $p$. New protected vertices arise whenever leafdegree-$1$ vertices are attached to. The attachment mechanism ensures that only leaves or neighbors of leaves, but not protected vertices, can be attached to. Thus $dP/dN$ has no loss term, and in particular,

\begin{equation}
\frac{d P}{dN}=\frac{M_1}{2N_1}\simeq \frac{M_1}{N},
\end{equation}
resulting in the unusual equality $m_1=p$. Figure \ref{fig:emb} depicts a sample from the model, highlighting protected vertices and leafdegree-$1$ vertices. The equal asymptotic abundance of these types of vertices, $m_1=p$, is a peculiar feature that does not occur for other models where the fraction of protected vertices is known \cite{Bona14,Devroye14,Mahmoud15,Janson15,Pittel17}.

We next consider $M_\ell$ for $\ell\ge 1$, taking care at $\ell=1$ because $M_1$ grows by one whenever a leaf is attached to (probability $1/2$), but increases by two whenever a leafdegree-$2$ vertex has one of its leaves attached to. In turn, $M_1$ decreases whenever direct or leaf attachment to a vertex with $\ell=1$ occurs. Thus 
\begin{equation}
\frac{dM_1}{dN}=\frac{M_2-M_1}{N_1}+\frac{1}{2},
\end{equation}
leading to $m_1=2m_2-2 m_1+\frac{1}{2}$, 
which together with $m_1=p$ give 
\begin{equation}
m_2=\frac{3p}{2}-\frac{1}{4}.
\end{equation}
The rates of change of $M_\ell$ for $\ell\ge 2$ follow a common pattern: $M_\ell$ grows by attachment to a vertex with leafdegree $\ell-1$ or by attachment to a leaf whose neighbor has leafdegree $\ell+1$. Similarly, $M_\ell$ decreases in case of either direct attachment or leaf attachment to a leafdegree-$\ell$ vertex. In the $N\rightarrow\infty$ limit this leads to
\begin{equation}
\label{eq:recurrence}
(\ell+1)m_{\ell+1}-(1+2\ell)m_\ell+(\ell-1)m_{\ell-1}=0, \ \ell\ge 2.
\end{equation}

Altogether, 
$(m_\ell)_{\ell\ge 0}$ thus satisfy
\begin{equation}
\begin{aligned}
\label{eq:ml_of_p_cr}
m_0&=p+\frac{1}{2},\quad m_1=p,\quad m_2=\frac{3p}{2}-\frac{1}{4},\\
m_{\ell}&=\left(2-\frac{1}{\ell}\right)m_{\ell-1}-\left(1-\frac{2}{\ell}\right)m_{\ell-2}, \  \ \ell\ge 3.
\end{aligned}
\end{equation}
We obtain $p$ below from a generating function (Sec.~\ref{app:crmodel_gf}). We note that $p$ is determined uniquely by normalization $\sum_{\ell\ge 0}m_\ell(p)=1$, and that $p$ can be numerically approximated to arbitrary precision by solving a truncated version of the normalization condition. Equations~\eqref{eq:ml_of_p_cr} can be combined (setting $m_{-1}=0$) to obtain a recurrence with no explicit dependence on $p$:
\begin{equation}
\begin{aligned}
\label{eq:mlcr}
(1+2\ell)m_\ell&=(\ell+1)m_{\ell+1}\\
&+(\ell-1)m_{\ell-1}+\frac{\delta_{\ell,1}+\delta_{\ell,0}}{2}, \ \ell\ge0.
\end{aligned}
\end{equation}
Letting $m(z)=\sum_{\ell\ge 0}m_\ell z^\ell$, Eq.~\eqref{eq:mlcr} is recast as 
\begin{equation}\begin{aligned}
m(z)+2\sum_{\ell} \ell m_\ell z^\ell &=\sum_{\ell}(\ell+1)m_{\ell+1}z^{\ell}\\
&+\sum_{\ell}(\ell-1)m_{\ell-1}z^\ell+\frac{1+z}{2}
\end{aligned}\end{equation}
from which we obtain an ODE
\begin{equation}
\label{eq:ode_cr}
(1-z)^2\frac{d m(z)}{dz}-m(z)+\frac{1+z}{2}=0.
\end{equation}

We analyze Eq.~\eqref{eq:ode_cr} in Sec.~\ref{app:crmodel_gf} to obtain the leafdegree distribution $(m_{\ell})_{\ell\ge 0}$ and protected fraction $p$. Since the attachment process depends on leafdegree rather than degree, the degree distribution $(n_{k})_{k\ge 1}$ does not satisfy a closed equation.  However, using the main approach described in Sec.~\ref{results}, we can rely on the joint degree-leafdegree distribution $n_{k,\ell}$ which indeed satisfies a closed recurrence
\begin{eqnarray}
\label{rec:LPA-0}
(2\ell+1)n_{k,\ell} &=& (\ell-1)n_{k-1,\ell-1}+(\ell+1)n_{k,\ell+1} \nonumber \\
&+&\frac{1}{2}(\delta_{k,1}\delta_{\ell,0} + \delta_{k,2}\delta_{\ell,1}).
\end{eqnarray}
The degrees vary in the range $0\leq \ell <k$. The number $N_{k,\ell}$ of vertices with degree $k$ and leafdegree $\ell$ is a random quantity, but we conjecture extensivity and asymptotic self-averaging. This implies convergence in probability of the form $\frac{1}{N}N_{k,\ell}\rightarrow n_{k,\ell}$ as $N\rightarrow\infty$, behavior that underlies our analysis. A stronger conjecture is the normality of random variables $\{N_{k,\ell}\}$, at least in the range $0\le \ell<k<N^{1/15}$ \cite{https://doi.org/10.1002/rsa.1009} as has been shown for $(N_k)_{k\ge 1}$ in the RRT \cite{Janson05}. The possibility of determining the degree distribution in this model is akin to derivation of $m_\ell\sim \ell^{-3}$ in the PA tree (and the exact Eq.~\eqref{eq:ml_PA}) as provided in (Sec.~\ref{ssec:ml_nk_pa}); in both cases, the joint distribution $n_{k,\ell}$ satisfies a closed recurrence, yet only one of ($n_k$, $m_{\ell}$) does. (In contrast with the RRT, for which both $n_k$ and $m_\ell$ do satisfy closed recurrences \cite{hartle2025statistics}.)

\subsubsection{Leafdegree distribution}
\label{app:crmodel_gf}

Solving Eq.~\eqref{eq:ode_cr} subject to $m(1)=1$ yields
\begin{equation}
\begin{aligned}
\label{mz:sol-LPA-0}
m(z) &= \int_0^{1-z}\frac{dw}{2w^2}\,(2-w)\,\exp\!\left[\frac{1}{1-z}-\frac{1}{w}\right]\\
&= 1+ \frac{e^{\frac{1}{1-z}}}{2}\mathrm{Ei}\left(\frac{-1}{1-z}\right),\\
\end{aligned}
\end{equation}
where $\text{Ei}(x)=\int_{-\infty}^x \frac{e^t}{t}\,dt$ is the exponential integral. Specializing \eqref{mz:sol-LPA-0} to $z=0$ we obtain
\begin{equation}
m_0=
1+\frac{e \text{Ei}(-1)}{2}.
\end{equation}

As shown in Sec.~\ref{app:crmodel_recu}, the fraction of protected vertices equals the fraction of vertices with leafdegree one, so 
\begin{equation}
p = m_1 = 
\frac{1+e \text{Ei}(-1)}{2} = 0.201\,826\ldots.
\end{equation}

\begin{figure}
\includegraphics[scale=0.75,trim=20 20 0 0]{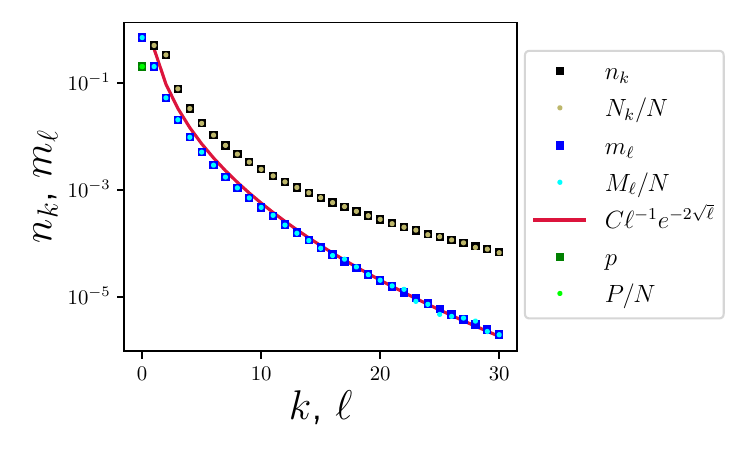}\caption{Leafdegree and degree distributions of the CR model at $\alpha=-\infty$ and $r=\frac{1}{2}$ as considered in Ref.~\cite{QPA}. The exact $m_\ell$ (blue squares), the numerical fractions $M_\ell/N$ in a single random tree of size $N=10^7$ (cyan circles), and the WKB asymptotic Eq.~\eqref{ml_cr} (red curve). The roughly estimated coefficient $C\approx 3.2$ is not analytically determined by the WKB approach. Additionally shown: the exact degree distribution $n_k$ (black squares) and numerical fractions $N_k/N$ (beige circles), and the exact and numerical protected fractions, $p$ and $P/N$ (dark green square and light green circle, respectively). }
\label{fig_ml_cr}
\end{figure}

Expanding \eqref{mz:sol-LPA-0} to higher powers of $z$ we find
\begin{equation}
\label{m-ell:LPA-0}
\begin{split}
&m_2 = \frac{1}{2}+\frac{3e \text{Ei}(-1)}{4}=0.052\,739\ldots\\
&m_3 = \frac{2}{3}+\frac{13e \text{Ei}(-1)}{12} = 0.020\,623\ldots\\
&m_4 = \frac{11}{12}+\frac{73e \text{Ei}(-1)}{48}= 0.009\,721\ldots\\
&m_5 = \frac{5}{4}+\frac{167e \text{Ei}(-1)}{80}=0.005\,124\ldots
\end{split}
\end{equation}
etc. Extracting the tail of the leafdegree distribution from the generating function \eqref{mz:sol-LPA-0} requires some effort, so we rely instead on a WKB approach  \cite{Bender} that allows us to deduce the tail up to a multiplicative constant. The form of the recurrence \eqref{eq:recurrence} suggests to seek the solution in the form $m_\ell= \ell^{-1}e^{-S(\ell)}$ when $\ell\gg 1$. Plugging this ansatz into  \eqref{eq:recurrence} we obtain
\begin{equation}
e^{S(\ell)-S(\ell+1)}+e^{S(\ell)-S(\ell-1)}=2+\ell^{-1}
\end{equation}
Expanding the exponents and then $S(\ell)-S(\ell\pm 1)$ in Taylor series, we obtain $\frac{dS}{d\ell} = \frac{1}{\sqrt{\ell}}+\frac{3}{8\ell^2}+\ldots$ from which $S=2\sqrt{\ell} + O(1)$. Thus
\begin{equation}
\label{ml_cr}
m_\ell \simeq  C\,\ell^{-1}\,e^{-2\sqrt{\ell}}
\end{equation}
The amplitude $C$ is not determined by the WKB approach. In Fig.~\ref{fig_ml_cr} the asymptotic Eq.~\eqref{ml_cr} is displayed alongside the exact values of $m_\ell$ and the empirical values of $M_\ell/N$ in a large tree sampled from the model.

\subsubsection{Moments of leafdegree distribution}

Consider the descending factorial defined by
\begin{equation}
\ell^{\underline{b}}=\prod_{k=0}^{b-1}(\ell-k)=\ell(\ell-1)\cdots(\ell-b+1).
\end{equation}
The associated descending moment of distribution $m_{\ell}$ is $\mu_b=\langle \ell^{\underline{b}}\rangle$. In terms of the generating function $m(z)$, the following identity holds: 
\begin{equation}
\label{mom-leaves}
\frac{d^b m}{d z^b}\Big|_{z=1}=
\langle \ell(\ell-1)\cdots(\ell-b+1)\rangle=\mu_b,
\end{equation}
allowing one to determine the descending moments of the leafdegree distribution. Rewriting Eq.~\eqref{mz:sol-LPA-0} as 
\begin{equation}
\label{mz:sol-1}
m(z) = 1-\frac{1-z}{2}\int_0^\infty\frac{du}{1+(1-z)u}\,e^{-u}
\end{equation}
simplifies the computation of the derivatives of the generating function at $z=1$. One finds 
\begin{equation}
\label{moments:d}
\mu_b = \langle \ell^{\underline{b}}\rangle = 
 \frac{b!(b-1)!}{2}
\end{equation}
valid for all integer $b\geq 0$. The standard moments $\langle \ell^j\rangle$ can be deduced from the descending moments. The first three standard moments are
\begin{equation}
\label{moments}
\langle \ell \rangle =  \frac{1}{2}\,, \quad \langle \ell^2 \rangle =  \frac{3}{2}\,, \quad \langle \ell^3 \rangle =  \frac{19}{2}.
\end{equation}

Generally, the identity
\begin{equation}
\ell^b=\sum_{j=0}^{b}\left\{b \atop j\right\}\ell^{\underline{j}}
\end{equation}
 involving Stirling numbers $\left\{b \atop j\right\}$ of the second kind \cite{Knuth} allows us to express the standard moments via descending moments
\begin{equation}
 \langle \ell^b\rangle=\sum_{j=0}^{b}\left\{b \atop j\right\}\langle \ell^{\underline{j}}\rangle = 
 \frac{1}{2}\sum_{j=0}^{b}\left\{b \atop j\right\} j!(j-1)!
\end{equation}
One finds that the standard moments of the leafdegree distribution are odd integers divided by $2$:
\begin{equation}
\begin{aligned}
2\langle \ell^4\rangle&=231,\\
2\langle \ell^5\rangle&=4\, 651,\\
2\langle \ell^6\rangle&=140\, 103,\\
2\langle \ell^7\rangle&=5\, 900\, 539,\\
2\langle \ell^8\rangle&=331\, 082\, 391,\\
2\langle \ell^9\rangle&=23\, 873\, 276\, 971,\\
2\langle \ell^{10}\rangle&=2\, 151\, 075\, 108\, 903.
\end{aligned}
\end{equation}

\subsection{Joint distribution}
\label{ssec:jointcrmodel}

The recurrence \eqref{rec:LPA-0} is obtainable by incorporating degree information into the associated leafdegree recurrence, Eq.~\eqref{eq:mlcr}, noting that (i) direct attachments increase both leafdegree and degree, (ii) attachments to neighboring leaves decrease leafdegree, and (iii) that $k=2$ and $k=1$ correspond to the source terms at $\ell=1$ and $\ell=0$. It can also be deduced directly as in Ref.~\cite{QPA} via reasoning akin to that in Sec.~\ref{pa} and Appendix~\ref{rrt}. The recurrence \eqref{rec:LPA-0} is more challenging to analyze than the recurrences for the PA tree and RRT, Eqs.~\eqref{eq:recursion_pa} and \eqref{eq:recursion_rrt}. We therefore begin with some special cases.

\subsubsection{Small and large $\ell$}
The fractions $n_{k,k-1}$ of near-stars satisfy
\begin{equation}
\label{nkk-12}
(2k-1)n_{k,k-1} = (k-2)n_{k-1,k-2}
\end{equation}
when $k\geq 3$. Solving \eqref{nkk-12} subject to $n_{2,1}=\frac{1}{6}$ gives
\begin{equation}
n_{k,k-1} = \frac{(k-2)!}{2\,(2k-1)!!}\,, \qquad k\geq 2.
\end{equation}
As such, the total fraction of near-stars (including leaves, $n_{1,0}=\frac{1}{2}$) is
\begin{equation}
\sum_{k\ge 1}n_{k,k-1}=\frac{3}{2}-\frac{\pi}{4}=0.714\, 501\, 836\, \ldots
\end{equation}

The fractions $n_{k,k-2}$ satisfy
\begin{eqnarray*}
(2k-3)n_{k,k-2} &=& (k-3)n_{k-1,k-3}+(k-1)n_{k,k-1} \\
&=&(k-3)n_{k-1,k-3}+\frac{(k-1)!}{2\,(2k-1)!!}
\end{eqnarray*}
for $k\geq 4$. This recurrence admits an explicit solution; similarly for $n_{k,k-3}$, but the expressions are increasingly cumbersome.

In the opposite extreme, when the leafdegree is small, one can express $n_{k,\ell}$ via the degree-stratified fractions 
$n_{k,0}$ of protected vertices:
\begin{subequations}
\begin{align}
n_{k,1} &=n_{k,0},
                                             \    \ \qquad\qquad\qquad ~~(k\geq 2),\\
n_{k,2} &=\tfrac{3}{2}n_{k,0},
                         \     \ \qquad\qquad\qquad (k\geq 3) ,\\
n_{k,3} &=\tfrac{5}{2}n_{k,0}-
\tfrac{1}{3}n_{k-1,0},
 \qquad\, (k\geq 4) ,          \\
n_{k,4} &=\tfrac{35}{8}n_{k,0}
- \tfrac{4}{3}n_{k-1,0},
\quad~~ \, (k\geq 5).
\end{align}
\end{subequations}
etc.  The first relation, $n_{k,1}=n_{k,0}$, can be summed to recover the unusual relation $m_1=p$, but is a much stronger statement: at any value of degree $k$, the protected fraction equals the leafdegree-$1$ fraction.

\subsubsection{Exact joint distribution values from recurrence}
\label{ssec:crmodel_nkl}

One can determine the exact values of $n_{k,\ell}$ directly from the recurrence \eqref{rec:LPA-0}. The values of $n_{k,\ell}$ with $k\leq 6$ are presented in Table \ref{tab_nkl}. To efficiently construct $n_{k,\ell}$ for all $0\le \ell<k\le k^*$ for some chosen maximum index $k^*$, we use the fact that in Eq.~\eqref{rec:LPA-0}, $n_{k,\ell}$ depends only on $n_{k-1,\ell-1}$, $n_{k,\ell+1}$, and $\ell$. Taking $n_{1,0}=\frac{1}{2}$ and $n_{2,0}=n_{2,1}=\frac{1}{6}$ as given, the algorithm proceeds to evaluate entries $n_{k,\ell}$ in ascending order of $k=3,\ldots,k^*$, and at each $k$, traversing the values of $\ell$ in descending order ($\ell=k-1,k-2,\ldots,0$), enforcing boundary conditions $n_{k,k}=n_{k,-1}=0$. In particular, the order of evaluation is $n_{3,2}$, $n_{3,1}$, $n_{3,0}$, $n_{4,3}$, $n_{4,2}$, \ldots, $n_{k^*,0}$. This ordering ensures that when computing $n_{k,\ell}$, both $n_{k-1,\ell-1}$ and $n_{k,\ell+1}$ have already been computed (or are boundary terms).

 Using the values of $n_{k,\ell}$, the exact degree distribution $n_k$ is obtainable by summation. See Table \ref{Table:degree} for $n_k$ at $k\le 7$. The values of $n_k$ align well with the fractions $N_k/N$ in a large random tree generated from the model; see Fig.~\ref{fig_ml_cr}. Visually, $n_k$ appears compatible with a stretched exponential decay akin to Eq.~\eqref{ml_cr}.

\begin{table}[h!]
\centering
\renewcommand{\arraystretch}{1.5}{
\begin{tabular}{| c | c | c | c | c | c | c |}
\hline
\diagbox{$k$}{$\ell$}  & $0$ & $1$ & $2$ & $3$ & $4$ & $5$\\ 
\hline
$1$ & $\frac{1}{2}$  &   \    &  \    & \      & \   & \   \\ 
\hline
$2$ & $\frac{1}{6}$   & $\frac{1}{6}$ &  \    &  \       &  \  & \   \\
\hline
$3$ & $\frac{1}{45}$  & $\frac{1}{45}$ &  $\frac{1}{30}$ &\  & \   & \   \\
\hline
$4$ & $\frac{32}{4725}$ & $\frac{32}{4725}$ & $\frac{16}{1575}$ & $\frac{1}{105}$     & \   & \    \\
\hline
$5$ & $\frac{1384}{496125}$ & $\frac{1384}{496125}$ & $\frac{692}{165375}$ & $\frac{52}{11025}$ &$\frac{1}{315}$  & \   \\
\hline 
$6$ & $\frac{777088}{573024375}$ & $\frac{777088}{573024375}$ & $\frac{388544}{191008125}$ & $\frac{93992}{38201625}$ &$\frac{2416}{1091475}$  & $\frac{4}{3465}$   \\
\hline 
\end{tabular}
}
\caption{The values $n_{k,\ell}$ with $k=1,\ldots,6$ and $\ell=0,\ldots,k-1$. Each $n_{k,\ell}$ was determined by direct solution of the two-index recurrence Eq.~\eqref{rec:LPA-0} (see Sec.~\ref{ssec:crmodel_nkl}).}
\label{tab_nkl}
\end{table}

\begin{table}[h!]
\centering
\renewcommand{\arraystretch}{1.5}{
\begin{tabular}{| c | c | c | c | c | c | c | c | c | c |}
\hline
$k$                         & 1                    & 2                           &  3                          & 4                                     &  5               & 6      & 7  \\ 
\hline 
$n_k$   & $\frac{1}{2}$   &   $\frac{1}{3}$   &  $\frac{7}{90}$ &  $\frac{157}{4725}$   &$\frac{8759}{496125}$  &$\frac{6059588}{573024375}$ & $\frac{19626289636}{ 2867986996875 }$\\ 
\hline
\end{tabular}
}
\caption{The degree distribution $n_k$ for $k \leq 7$, obtained as $n_k=\sum_{\ell=0}^{k-1}n_{k,\ell}$, with $n_{k,\ell}$ computed exactly as per Sec.~\ref{ssec:crmodel_nkl}.} 
\label{Table:degree}
\end{table}

\subsubsection{Bivariate generating function}
\label{app:crmodel_joint}

Using the generating function 
\begin{equation}
g(y,z) =\sum_{0\le \ell<k<\infty}y^k z^\ell n_{k,\ell},
\end{equation}
we recast the recurrence \eqref{rec:LPA-0} into 
\begin{equation}
\label{pde:LPA-0}
(1-2z+yz^2)\,\frac{\partial g}{\partial z}-g+\frac{y+y^2z}{2}=0.
\end{equation}
The factor $1-2z+yz^2$ is negative in the interval
\begin{equation}
\label{z-star}
z_* < z <1, \qquad  z_*= \frac{1-\sqrt{1-y}}{y}.
\end{equation}
In this region, the solution reads 
\begin{eqnarray}
\label{g:sol-LPA-0}
g = -\int_{z_*}^z dw\,\frac{y+y^2 w}{2(1-2w + y w^2)}\,e^{F(y,z)-F(y,w)},
\end{eqnarray}
where we shortly write
\begin{equation}
F(y,z)  = \frac{1}{\sqrt{1-y}}\,\tanh^{-1}\!\left(\frac{1-y z}{\sqrt{1-y}}\right).
\end{equation}

{\it Degree distribution.} Specializing to $z=1$ we obtain
\begin{equation}
\label{rec:nk}
\sum_{k\geq 1} y^k n_k=-\frac{e^{F(y,1)}}{2}\int_{z_*}^1 dw\,\frac{y+y^2 w}{1-2w + y w^2}\,e^{-F(y,w)},
\end{equation}
with
\begin{equation}
\label{f-y}
e^{F(y,1)}=\exp\!\left[\frac{\tanh^{-1}\sqrt{1-y}}{\sqrt{1-y}}\right].
\end{equation}
Expanding the right-hand side of \eqref{rec:nk} and finding an explicit general formula for the degree distribution appears impossible.

{\it Protected degree distribution.} We recall that $(n_{k,0})_{k\ge 1}$ are generated by $g(y,0)$. Thus taking the $z=0$ limit of Eq.~\eqref{g:sol-LPA-0} we have
\begin{equation}
\label{gy0_cr}
\sum_{k\geq 1} y^k n_{k,0}= \frac{e^{F(y,0)}}{2}\int^{z_*}_0 dw\,\frac{y+y^2 w}{1-2w + y w^2}\,e^{-F(y,w)},
\end{equation}
with
\begin{equation}
F(y,0)= \frac{1}{\sqrt{1-y}}\,\tanh^{-1}\!\left(\frac{1}{\sqrt{1-y}}\right).
\end{equation}
Similarly to Eq.~\eqref{rec:nk}, obtaining coefficients $n_{k,0}$ from Eq.~\eqref{gy0_cr} seems intractable.

\subsubsection{Descending moments of joint distribution}

One would also like to compute the moments of the degree distribution, and the mixed moments involving the degree and leafdegree distributions. These moments can be found recursively as we demonstrate below. One can also deduce general formulae resembling the moments \eqref{moments:d} of the leafdegree if the degree is fixed. 

To determine the moments, we use the generating function \eqref{eq:gdef}) and compute the moments using the governing equation \eqref{pde:LPA-0} for the generating function, thus circumventing the usage of an intractable exact solution (Eq.~\eqref{g:sol-LPA-0}).  The computation relies on the identities
\begin{equation*}
\langle k^{\underline{a}}\,\ell^{\underline{b}}\rangle = \partial_y^a\partial_z^b g|_{y=z=1},
\end{equation*}
valid for all integer $a\geq 0$ and $b\geq 0$. When $a=0$, these identities coincide with \eqref{mom-leaves}. Note, application of $\partial_y$ to Eq.~\eqref{pde:LPA-0} and setting $y=z=1$ recovers $\langle k\rangle=2$.
The quadratic descending moments are
\begin{equation}
\label{k-ell:2}
\langle k(k-1)\rangle=7, \quad \langle k\ell\rangle = 3, \quad \langle \ell(\ell-1)\rangle=1.
\end{equation}

Similar calculations give ternary descending moments 
\begin{equation}
\langle k^{\underline{3}}\rangle= 135,\quad \langle k^{\underline{2}}\,\ell\rangle=45,\quad \langle k\,\ell^{\underline{2}}\rangle=17,\quad 
\langle \ell^{\underline{3}}\rangle= 6
\end{equation}
and quartic descending moments 
\begin{equation}
\begin{split}
\langle k(k-1)(k-2)(k-3)\rangle =\langle k^{\underline{4}}\rangle&= 9048,\\
\langle k(k-1)(k-2)\ell\rangle =\langle k^{\underline{3}}\,\ell\rangle&= 2262,\\
\langle k(k-1)\ell(\ell-1)\rangle =\langle k^{\underline{2}}\,\ell^{\underline{2}}\rangle&= 664,\\
\langle k\ell(\ell-1)(\ell-2)\rangle=\langle k\,\ell^{\underline{3}}\rangle &= 216,\\
 \langle \ell(\ell-1)(\ell-2)(\ell-3)\rangle = \langle\ell^{\underline{4}}\rangle&=72.
\end{split}
\end{equation}

\subsubsection{Recurrences for descending moments}

Applying $\partial_z^b$ to Eq.~\eqref{pde:LPA-0} and setting $y=z=1$ one arrives at a recurrence which is solved to recover already known familiy of moments \eqref{moments:d}. To determine 
\begin{equation*}
\nu_b=\langle k\ell^{\underline{b}}\rangle= \langle k\, \ell\cdots(\ell-b+1)\rangle = \partial_y\partial_z^b g|_{y=z=1},
\end{equation*}
we apply $\partial_y\partial_z^b$ to Eq.~\eqref{pde:LPA-0}, set $y=z=1$, and arrive at a inhomogeneous recurrence
\begin{equation}
\label{nu-b}
\nu_b = b(b-1)\nu_{b-1}+b(b-1)\mu_{b-1}+2b\mu_b+\mu_{b+1},
\end{equation}
for $b\geq 2$. Recalling \eqref{moments:d}, we simplify \eqref{nu-b} to 
\begin{equation}
\nu_b = b(b-1)\nu_{b-1}+\frac{ (1+3b+b^2)(b-1)! b!}{2}
\end{equation}
for $b\geq 2$ which we should solve subject to the `boundary' condition $\nu_1=\langle k\ell\rangle = 3$. The solution reads
\begin{equation}
\label{nu-b-sol}
\nu_b=  \frac{(b^2+5b+3)(b-1)!(b+1)!}{6}.
\end{equation}

The following family of moments 
\begin{equation*}
\lambda_b= \langle k^{\underline{2}}\ell^{\underline{b}}\rangle= \langle k(k-1)\, \ell\cdots(\ell-b+1)\rangle = \partial_y^2\partial_z^b g|_{y=z=1}
\end{equation*}
can be subsequently determined. Similarly to \eqref{nu-b} one finds
\begin{equation}
\label{lambda-b}
\lambda_b = b(b-1)\lambda_{b-1}+2b(b-1)\nu_{b-1}+4b\nu_b+2\nu_{b+1}.
\end{equation}
Combining this equation with \eqref{nu-b-sol} we obtain
\begin{equation*}
\lambda_b = b(b-1)\lambda_{b-1}+\frac{ (b^4+12 b^3+45 b^2+60 b+23)(b!)^2}{3},
\end{equation*}
which is solved subject to $\lambda_1=\langle k(k-1)\ell\rangle = 45$ to yield
\begin{equation}
\label{lambda-b-sol}
\lambda_b =  \frac{\Lambda(b)}{90}\,(b-1)!b!,
\end{equation}
where we shortly write $\Lambda(b)$ for the polynomial 
\begin{equation*}
5 b^6+87 b^5+530 b^4+1395 b^3+1580 b^2+633 b-180.
\end{equation*}

\end{document}